\documentclass[pre,epsfig,address,twocolumn]{revtex4}%
\usepackage[english]{babel}
\usepackage[latin1]{inputenc}
\usepackage[dvips]{graphicx}
\usepackage{amsmath}
\usepackage{amssymb}
\usepackage{epsfig}
\usepackage{array}
\usepackage{amsfonts}
\usepackage{graphicx}
\usepackage{subfigure}
\usepackage{color}%
\begin{document}
\title{ Hierarchy of boundary driven phase transitions in multi-species particle systems}
\author{Vladislav Popkov$^{1}$ and Mario Salerno$^{1}$ }
\affiliation{$^{1}$ Dipartimento di Fisica "E.R. Caianiello", and Consorzio Nazionale
Interuniversitario per le Scienze Fisiche della Materia (CNISM), Universit\`a
di Salerno, Fisciano, Italy}
\date{\today }

\begin{abstract}
Interacting systems with $K$ driven particle species on a open chain or chains
which are coupled at the ends to boundary reservoirs with fixed particle
densities are considered. We classify discontinuous and continuous phase
transitions which are driven by adiabatic change of boundary conditions. We
build minimal paths along which any given boundary driven phase transition
(BDPT) is observed and reveal kinetic mechanisms governing these transitions.
Combining minimal paths, we can drive the system from a stationary state with
all positive characteristic speeds to a state with all negative characteristic
speeds, by means of adiabatic changes of the boundary conditions. We show that
along such composite paths one generically encounters $Z$ discontinuous and
$2(K-Z)$ continuous BDPTs with $Z$ taking values $0\leq Z\leq K$ depending on
the path. As model examples we consider solvable exclusion processes with
product measure states and $K=1,2,3$ particle species and a non-solvable
two-way traffic model. Our findings are confirmed by numerical integration of
hydrodynamic limit equations and by Monte Carlo simulations. Results extend
straightforwardly to a wide class of driven diffusive systems with several
conserved particle species.

\end{abstract}

%\pacs{64.60.Cn, 05.70.Ln, 02.50.Ga}
\maketitle

\section{Introduction}

\label{sec::Introduction}

Systems of interacting particles out of equilibrium with more than one
particle species find wide range of applications in biological, social and
physical context \cite{Gunter03TwoSpecies_review}, \cite{KoloFischer07_review}%
, \cite{Frey09}. Remarkable phenomena occurring in these systems are the
so-called boundary-driven phase transitions (BDPTs), e.g. phase transitions
induced uniquely by changes of the boundary conditions \cite{Krug91}. These
transitions can be both of first and second order, depending on whether the
order parameter (e.g. the stationary bulk density) changing discontinuously or
continuously across the transition line. It has been demonstrated that in
systems with one particle species the first order (discontinuous) BDPTs are
governed by shocks dynamics, while the second order (continuous) BDPTs are
governed by rarefaction waves \cite{Kolo98}. In case of several particle
species corresponding in the hydrodynamic limit to the hyperbolic systems of
conservation laws, it was noted that the number of qualitatively different
first-order BDPTs increases with the number of particle species
\cite{PopkovCambridge}, a fact which is related to a complex structure of
shocks in systems of conservation laws \cite{Lax73,Lax2006},\cite{Bressan}.
However, generic properties of boundary driven phase transitions in
multi-species driven particle systems are largely unknown. Also, the
mechanisms governing stationary state selection are known for one-species
systems \cite{Kolo98,Gunter_Slava_Europhys}, while for systems with several
particle species such a knowledge is lacking.

The aim of the present paper is to show the existence of hierarchies of
qualitatively different phase transitions in driven systems with several
particle species, with open boundaries, and to provide a first classification
of them. In particular we show that in systems with $K$ species of particles
there are generically at least $K$ first order (discontinuous) and $K$ second
order (continuous) qualitatively different phase transitions. We demonstrate
that these phase transitions can be observed by driving the system from a
stationary state with all positive characteristic velocities to a state with
all negative characteristic velocities, by means of adiabatic changes of the
densities at the boundaries. The kinetic mechanisms underlying the occurrence
of first order BDPTs is ascribed to shock waves interactions (both among
themselves and with boundaries) while the origin of second order phase
transitions is shown to be connected with rarefaction waves. This leads to
several physical implications which are confirmed by numerical simulations. In
particular, we show that the existence of shock waves and the continuity of
the flux across a first order BDPT allows us to predict the location of the
stationary densities after their discontinuous change. Similarly, conditions
of stability of rarefaction waves allows us to predict the location of the
second order phase transition points along a specific path in parameter space.
Also, it follows that the occurence of qualitatively different BDPTs is
connected to the existence of different classes of shock waves and rarefaction
waves in systems with several particle species. Continuous paths in parameter
space along which sequences of $Z$ first order and $2(K-Z)$ second order
transitions with ($Z$ $=0,1,...,K$) occur, are identified. We show that for
each value of $Z$, there are $\binom{K}{Z}$ qualitatively different paths,
each of them leading to\ a different set of transitions.

Our general approach is tested on ideal solvable models (e.g. which admit
product steady states) as well as on more realistic (not exactly solvable)
ones. For the former we derive hydrodynamic equations of motion and use the
theory of shock and rarefaction waves for system of conservation laws for
their investigation. For non-solvable models we recourse to Monte-Carlo
simulations as principal tool of investigation. To simplify the presentation
we concentrate in more detail on the case of two particle species but the
results are discussed in a manner that the extension to the case of an
arbitrary number of particle species becomes straightforward.

The plan of the paper is the following. In Section
\ref{sec::Multi-species particle models out of equilibrium} we introduce
multi-species particle models and their basic properties. In Sec.
\ref{sec::Hierarchy of continuous and discontinuous phase transitions} we
discuss hierarchies of BDPTs in multi-species systems and in Sec.
\ref{sec::A minimal path} we characterize the "minimal path" along which any
chosen (discontinuous or continuous) BDPT can be observed, taking as working
examples the cases of one and two particle species ($K=1,2$). In Sec.s
\ref{sec::Shock waves interaction as a mechanism of the phase transitions of the first order}
and \ref{sec::Rarefaction waves govern continuous phase transitions} we
discuss the basic mechanisms governing first and second order BDPTs in
multi-species systems, respectively. In Sec. \ref{sec::Two way traffic model}
we illustrate our approach for the case of a more realistic (not exactly
solvable) model for bidirectional traffic, while in Sec.
\ref{sec::Special paths} we discuss special paths for models with $K$ particle
species along which sequences of BDPTs in the parameter space occur. Finally,
Sec.\ref{sec::Conclusions} serves for conclusions and for open problems.

\section{ Multi-species particle models out of equilibrium}

\label{sec::Multi-species particle models out of equilibrium}

We consider a system, consisting of many interacting particles, dynamics of
which is governed by a Master Equation, with the rates $\{\Gamma_{CC^{\prime}%
}\}$ of transition between the states $C$ and $C^{\prime}$, on a discrete
lattice. We assume that there are $K\geq1$ particle species conserved
independently, and that one can construct a hydrodynamic limit, which has the
form of a system of conservation laws, of the type%
\begin{align}
{\frac{\partial\rho_{k}}{\partial t}}+{\frac{\partial j_{k}(\rho_{1},\rho
_{2},...\rho_{K})}{\partial x}}  &  =\varepsilon{\frac{\partial}{\partial x}%
}\left(  \sum\limits_{k=1}^{K}B_{kj}\frac{\partial\rho_{j}}{\partial x}\right)
\label{PDE}\\
{k}  &  {=1,2,...,K.}\nonumber
\end{align}
Here $\rho_{k}\left(  x,t\right)  $ denotes the averaged local density of the
specie $k$, $j_{k}$ is the respective flux, $B_{ij}(\rho_{1},\rho_{2}%
,...,\rho_{K})$ is the diffusion matrix, $\varepsilon$ is an infinitesimally
small positive constant. The system is defined on a segment $x\in\lbrack0,1]$.
At the ends of the segment, Dirichlet boundary conditions are imposed:
\begin{equation}
\rho_{k}[0]=\rho_{k,L},\text{ \ }\rho_{k}[1]=\rho_{k,R}.
\label{PDEBoundaryConditions}%
\end{equation}
We are interested in the stationary state of the system, i.e. in a state
attained in the infinite time limit $t\rightarrow\infty$. \ Note that in the
case of perfect match between left and right boundary densities Eqs.
(\ref{PDE}),(\ref{PDEBoundaryConditions}) allow stationary space-homogeneous
solution $\rho_{k}(x)\equiv\rho_{k,L}=\rho_{k,R}$ for all $k$. Stability of
this solution for all physical values of $\rho_{1},\rho_{2},...,\rho_{K}$ is
guaranteed by the positive definiteness of the diffusion matrix $B_{ij}$. For
the source particle model, this corresponds to an existence of a homogeneous
stationary state with constant average particle densities $\rho_{1},\rho
_{2},...,\rho_{K}$.

Some remarks are in order. Dirichlet boundary conditions
(\ref{PDEBoundaryConditions}) in terms of particle system mean that the system
is coupled to boundary reservoirs with fixed particles densities $\rho_{k,L},
\rho_{k,R}$ at the left and right ends, respectively. How this can be
implemented or read off from boundary rates was discussed in
\cite{reflections_ABO}. The flux $j_{k}(\rho_{1},\rho_{2},...,\rho_{K})$ is a
nonlinear function of the particle densities, which implies interaction
between particles. For a precise definition of the genuine flux nonlinearity
see e.g. \cite{Lax73}.

A special role is played by the Jacobian matrix of the flux ${\left(
\mathcal{D}\mathbf{j}\right)  }_{kw}=\partial j_{k}/\partial\rho_{w}$ whose
eigenvalues $c_{1}<c_{2}<...<c_{K}$, called characteristic speeds, are all
real and distinct \cite{Weak_Hyperbolicity} (note that we numerate the
characteristic speeds in the increasing order). The characteristic speeds for
the original particle system are velocities with which the local perturbations
of a homogeneous state are propagating \cite{GunterSlava_StatPhys}. The
physical region (i.e. the region with physically meaningful particle densities
$\rho_{k}$) splits into domains according to the signs of characteristic
velocities. E.g. for $K=2$ we shall call $G_{-+}$ the domain in $u,v$ space
($u,v$ are particle densities of the two species) where $c_{1}(u,v)<0$ and
$c_{2}(u,v)>0$. Characteristic speeds are smoothly-changing functions of
particle densities. Special role is played by the subdomains of dimension
$K-1$ across which, one of characteristic speeds changes its sign. E.g. the
domains $G_{++}$and $G_{-+}$ are separated by the subdomain $G_{0+}$ with
$c_{1}(u,v)=0$. An example of a decomposition of the physical region for $K=2$
is given in Fig.\ref{Fig_Gdecomposition}. If a stationary state of a particle
model has bulk particle densities $u_{stat},v_{stat}$ belonging to the
$G_{++}$ region, we call it $G_{++}$ stationary state, etc.. Analogously, if
the left boundary densities $u_{L},v_{L}$ belong to, say, $G_{-+}$ domain, we
say that left boundary is of $G_{-+}$ type etc. The generalization to
arbitrary number of particle species $K>2$ is obvious.

\section{Hierarchies of continuous and discontinuous BDPT}

\label{sec::Hierarchy of continuous and discontinuous phase transitions}

Different types of BDPTs can occur in these models. A discontinuous (first
order) phase transition of the $p-th$ type is a transition between a
stationary state with $c_{1},...,c_{p-1}<0,$ $c_{p},...,c_{K}>0$ and a
stationary state where $c_{p}$ has changed its sign. E.g., for $K=2$, type $1$
and type $2$ discontinuous transitions are transitions between $G_{++}%
\leftrightarrows G_{-+}$ and $G_{-+}\leftrightarrows G_{--}$, respectively.
Note that there is no direct transition between the $G_{++}$ and $G_{--}$
state. This is due to the strict hyperbolicity: there is no region, where
$G_{++}$ and $G_{--}$ touch each other (the contrary would mean the existence
of a weak hyperbolic point \cite{Weak_Hyperbolicity}). Obviously, in systems
with $K$ species $p$ can generically take values $p=1,2,...K$, leading to $K$
different types of discontinuous phase transitions.

The continuous (second order) phase transition of type $p$ is a transition
between a stationary state with zero $p-th$ characteristic velocity, $c_{p}%
=0$, and a stationary state where $c_{p}$ is strictly positive or negative.
The signs of other characteristic velocities $c_{q},q\neq p$ do not change
across this transition. E.g. for $K=2$, transitions $G_{++}\leftrightarrows
G_{0+},G_{0+}\leftrightarrows G_{-+}$ are of type $p=1$, while the transitions
$G_{-+}\leftrightarrows G_{-0},G_{-0}\leftrightarrows G_{--}$ are of type
$p=2$. Note that continuous transition of type $G_{0+}\leftrightarrows G_{-0}$
does not happen since the respective subregions have no intersection due to
strict hyperbolicity. \begin{figure}[ptb]
\centerline{
\subfigure[\label{figMinPath:a}]{
\includegraphics[width=4.5cm,height=4.3cm,clip]{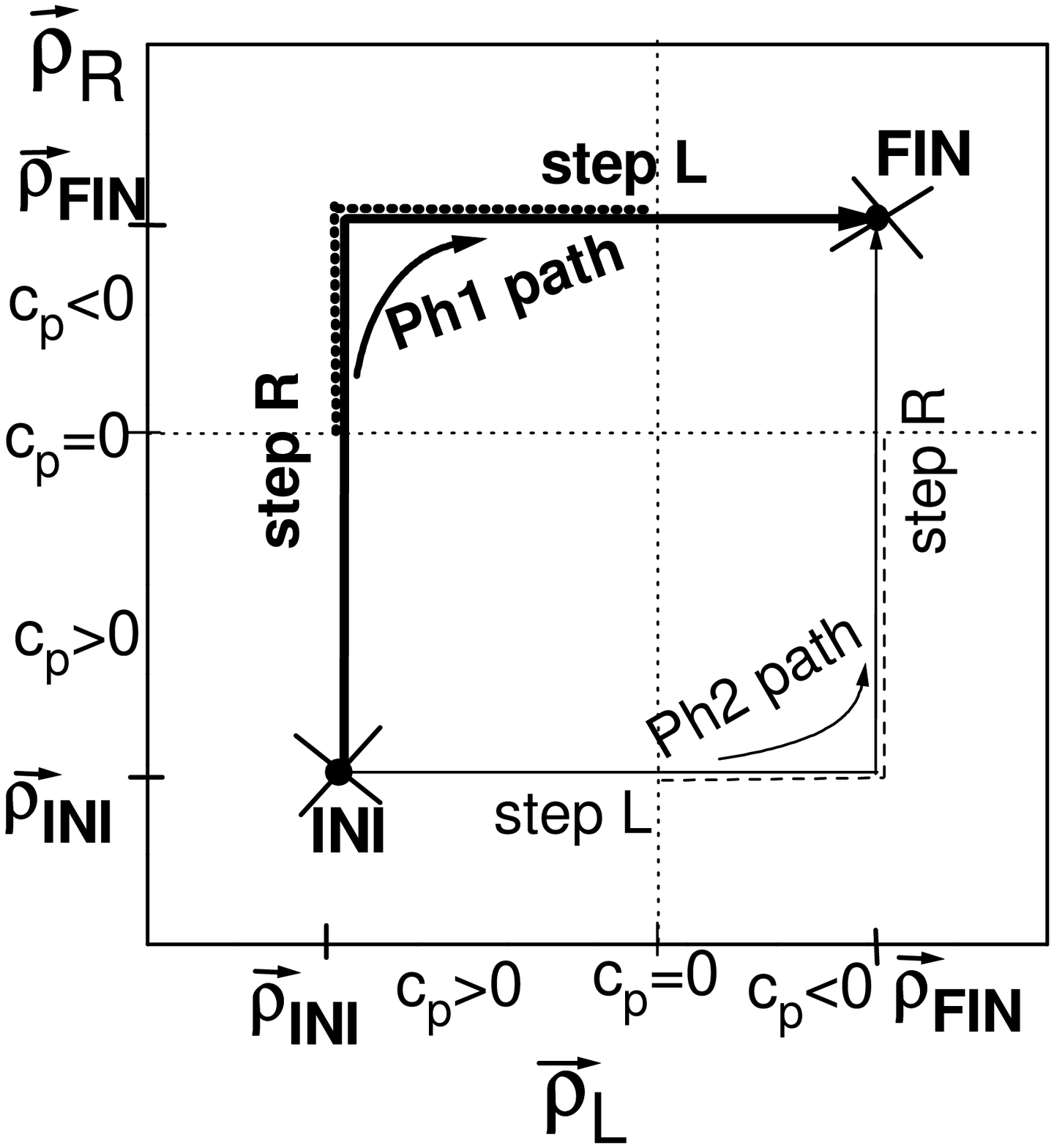}}
\subfigure[\label{figMinPath:b}]{
\includegraphics[width=4.6cm,height=4.2cm,clip]{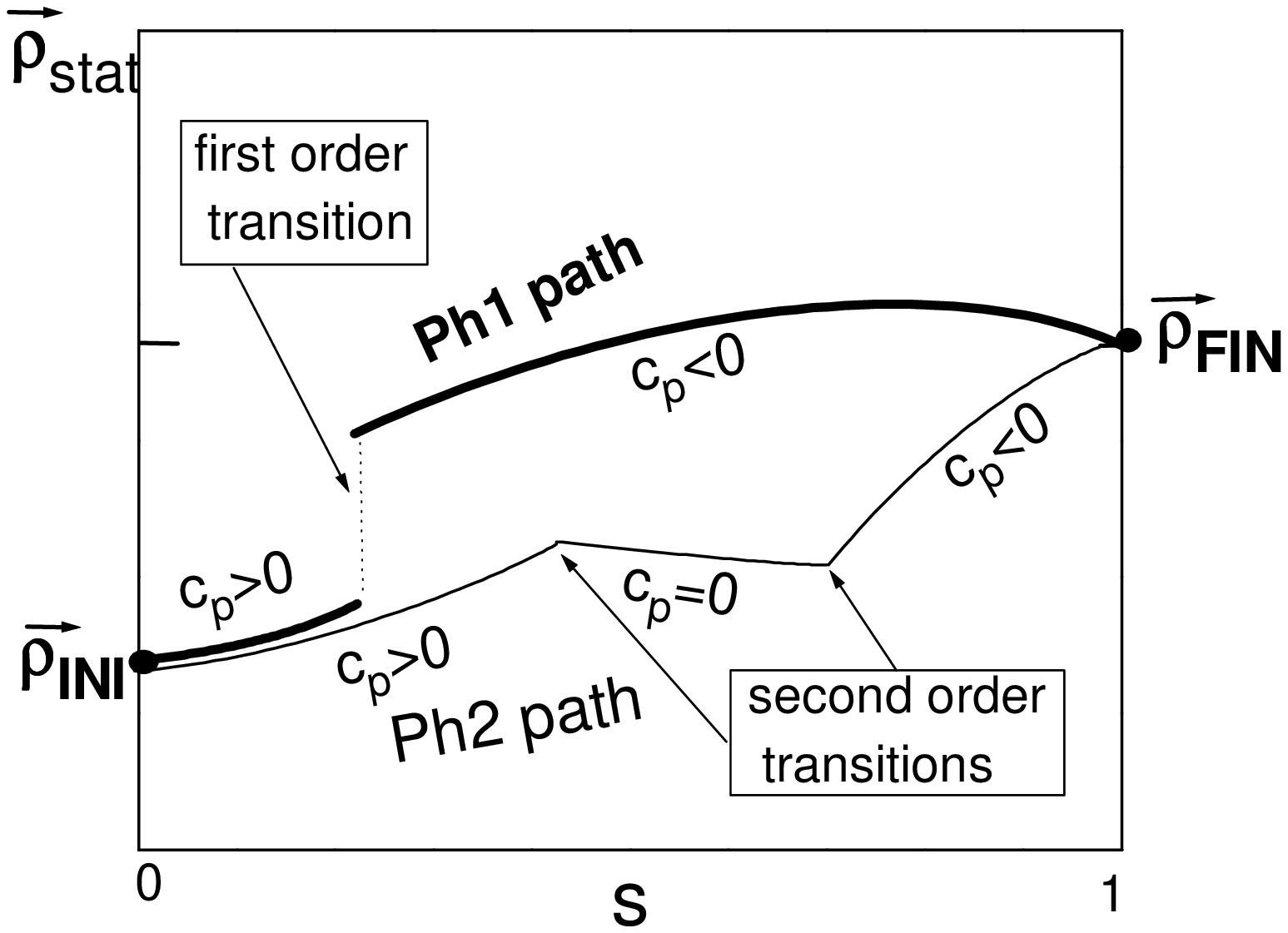}}
} \caption{A schematic diagram, showing minimal Ph1 path (bold lines) and Ph2
paths (thin lines) in parameter space (Panel (a)) and in physical region
(Panel (b)). The Ph2 path is continuous both in parameter space and in
physical region. The Ph1 path is continuous in parameter space but is
discontinuous in physical region. \textbf{Panel (a)}: The axes represent sets
of boundary densities at the left and at the right boundary. The part of the
Ph2 trajectory marked by dashed line indicate the rarefaction-wave governed
stationary state,(see
Sec.\ref{sec::Rarefaction waves govern continuous phase transitions}). The
bold dotted line in the left upper corner marks the part of the Ph1 path,
inside which a first order transition occurs, see
Sec.\ref{sec::Shock waves interaction as a mechanism of the phase transitions of the first order}%
. \textbf{Panel (b)}: Schematic evolution of the stationary densities along a
Ph1 path (bold broken line) and a Ph2 path (thin line). Along the Ph1 path at
least one discontinuous $p$-type phase transition is observed. Along the Ph2
path, a pinning and depinning to a state with zero characteristic velocity
(two continuous phase transitions of $p$-type) are observed. The pinning and
depinning point correspond to the ends of the dashed-marked Ph2 segment on
Panel (a). }%
\label{Fig_MinPath}%
\end{figure}In systems with weak hyperbolic point, where such an intersection
exists, a continuous transition $G_{0+}\leftrightarrows G_{-0}$ may become
possible (see direct Ph2 path in Fig.\ref{Fig_MC} (b)). In case of $K$
species, $p$ can generically take values $p=1,2,..K$, leading to $K$ different
types of continuous (second order) transitions.

\section{Minimal paths to observe BDPT}

\label{sec::A minimal path}

Here we describe the simplest path in parameter space along which a
\textit{single} first or second order phase transition is surely observed, in
a system with $K$ species of particles. For this we introduce a parameter
space of dimension $2K$, coordinates of which are given by the left and right
boundary densities $\{\rho_{L}|\rho_{R}\}\equiv\{\rho_{1,L},\rho
_{2,L},...,\rho_{K,L}|\rho_{1,R},...,\rho_{K,R}\}$. Each point of the
parameter space represents some boundary conditions
(\ref{PDEBoundaryConditions}). Let us also define a physical region of
dimension $K$ coordinates of which are average bulk particle densities
$\rho_{1},\rho_{2},...\rho_{K}$. Each stationary state with bulk particle
densities $\rho_{stat}\equiv\{\rho_{1}^{stat},\rho_{2}^{stat},...,\rho
_{K}^{stat}\}$ is represented by a point in the physical region. The bulk
stationary density will be our order parameter.

A path in parameter space is represented by a \textit{continuous} curve
$\Gamma(s)$ given by the left and right boundary densities $\{\rho_{L}%
(s)|\rho_{R}(s)\}$ along the path, parametrized by the running coordinate
$0\leq s\leq$ $1,$ with $s=0$ and $s=1$ corresponding to the initial and final
points of the path, respectively. For each value of $s$, we wait until the
system reaches a stationary state (which we assume to be homogeneous in the
bulk) and record the stationary bulk particle densities $\rho_{stat}%
(s)\equiv\{\rho_{1}^{stat}(s),\rho_{2}^{stat}(s),...,\rho_{K}^{stat}(s)\}$.
Thus, each path $\Gamma(s)$ in parameter space of dimension $2K$ is mapped on
a path $\rho_{stat}(s)$ in physical region of dimension $K$. The mapping
$\Gamma(s)\rightarrow\rho_{stat}(s)$ is not invertible, because many different
boundary conditions can lead to states with equal bulk stationary densities.

Consider two neighboring domains $G_{X}$ and $G_{Y}$ in the physical region
$\rho_{1},\rho_{2},...\rho_{K}$ characterized by the following signs of the
characteristic speeds $c_{i}(\rho_{1},\rho_{2},...\rho_{K})$:
\begin{align}
&  G_{X}:c_{1}<...<c_{p-1}<0;\text{ \ \ }c_{K}>c_{K-1}>...>c_{p}>0,\nonumber\\
&  G_{Y}:c_{1}<...<c_{p}<0;\text{ \ \ }c_{K}>c_{K-1}>...>c_{p+1}>0,
\label{G domains}%
\end{align}
Note that both $G_{X},G_{Y}$ have dimension $K$ and that domain $G_{X}$ has
one positive characteristic speed more than $G_{Y}$. We denote with $G_{X0Y}$
a subdomain of dimension $K-1$ separating domains $G_{X}$ and $G_{Y}$, given
by:
\begin{align}
&  G_{X0Y}:c_{1}<...<c_{p-1}<0;\text{ \ \ \ }c_{p}=0;\nonumber\\
&  \text{ \ \ \ \ \ \ }c_{K}>c_{K-1}>...>c_{p+1}>0.\nonumber
\end{align}

Phase transitions occur along paths which connect neigbouring domains. In the
following we denote by $\rho_{L}^{ini},\rho_{R}^{ini}$ the sets of the
particle densities in the left and right boundary reservoirs at the initial
point $s=0$ of a path $\Gamma(s)$, i.e. $\rho_{L}^{ini}\equiv\{\rho
_{k,L}(s=0)\}_{k=1}^{K}$ , \ $\rho_{R}^{ini}\equiv\{\rho_{k,R}(s=0)\}_{k=1}%
^{K}$, $\Gamma(s=0)\equiv\{\rho_{L}^{ini}|\rho_{R}^{ini}\}$. Analogously,
denote $\rho_{L}^{final},\rho_{R}^{final}$ the respective boundary densities
at the end of the path $\Gamma(s=1)\equiv\{\rho_{L}^{final}|\rho_{R}%
^{final}\}$. Let us choose the initial \ and the final boundary densities on
both boundaries from domains $G_{X}$ $\ $\ and $G_{Y}$, respectively,
\begin{equation}
\rho_{L}^{ini}\in G_{X},\rho_{R}^{ini}\in G_{X}\text{ \ and }\rho_{L}%
^{final}\in G_{Y},\rho_{R}^{final}\in G_{Y}. \label{RHO_ini and RHO_final}%
\end{equation}
In addition, we shall choose the initial and the final boundary densities
fully matching, $\rho_{L}^{ini}\equiv\rho_{R}^{ini}=\rho_{INI}$, and $\rho
_{L}^{final}\equiv\rho_{R}^{final}=\rho_{FIN}$. Since the full match allows
for trivial homogeneous stationary solution of (\ref{PDE}%
),(\ref{PDEBoundaryConditions}), the initial and final stationary states along
the path belong to different $G$ domains: $\rho_{stat}(s=0)=\rho_{INI}\in
G_{X}$ and $\rho_{stat}(s=1)=\rho_{FIN}\in G_{Y}$, see discussion after
Eq.(\ref{PDEBoundaryConditions}). Note that the condition of the full match
can be relaxed (for an example see Sec.\ref{sec::Two way traffic model}), and
is chosen here for simplicity of presentation.

Having chosen the initial and final points of a path $\Gamma(s)$, let us now
define paths $\Gamma^{Ph1}(s),$ $\Gamma^{Ph2}(s)$ of type Ph1 and Ph2 as
consisting of two consecutive elementary steps:%
\begin{align}
\Gamma^{Ph1}(s)  &  :\{\rho_{L}^{ini}|\rho_{R}^{ini}\}\rightarrow\{\rho
_{L}^{ini}|\rho_{R}^{final}\}\rightarrow\{\rho_{L}^{final}|\rho_{R}%
^{final}\},\label{Ph1}\\
\Gamma^{Ph2}(s)  &  :\{\rho_{L}^{ini}|\rho_{R}^{ini}\}\rightarrow\{\rho
_{L}^{final}|\rho_{R}^{ini}\}\rightarrow\{\rho_{L}^{final}|\rho_{R}^{final}\}.
\label{Ph2}%
\end{align}
During each elementary step the boundary densities at one boundary change
adiabatically while the other boundary is kept fixed. We shall call an
elementary step during which the left (the right) boundary densities change a
step L ( step R), respectively. Thus, a Ph1 path $\Gamma^{Ph1}(s)$ consists of
consecutive steps step R $\rightarrow$ step L, while in a Ph2 path the steps R
and L are interchanged, see Fig. \ref{Fig_MinPath}. It is important that a
domain $G_{X}$ of departure for both paths has more more positive
characteristic velocities than the domain of arrival $G_{Y}$, the reason for
which will be explained in the next section.

We assume trajectories $\rho_{R}^{ini}\longrightarrow\rho_{R}^{final}$ (step
R) and $\ \rho_{L}^{ini}\longrightarrow\rho_{L}^{final}$ (step L) to lie
entirely inside $G_{X}$ and $G_{Y}$, and to cross\ $c_{p}=0$ hyperplane only
once, see Fig.\ref{Fig_MinPath} (a). The main results of this section are
presented in the two propositions listed below.

\textbf{Proposition I}. \textit{Discontinuous change of the stationary density
}$\rho_{stat}$\textit{ from a point in domain }$G_{X}$ \textit{to a point in
neighboring domain }$G_{Y}$\textit{ (}$p-$\textit{th type of first order
transition) occurs along \textbf{any} Ph1 path (\ref{Ph1}) connecting these
domains.} \begin{figure}[ptb]
\centerline{
\subfigure[\label{ASEPPhase:a}]{
\includegraphics[width=4.5cm,height=4.5cm,clip]{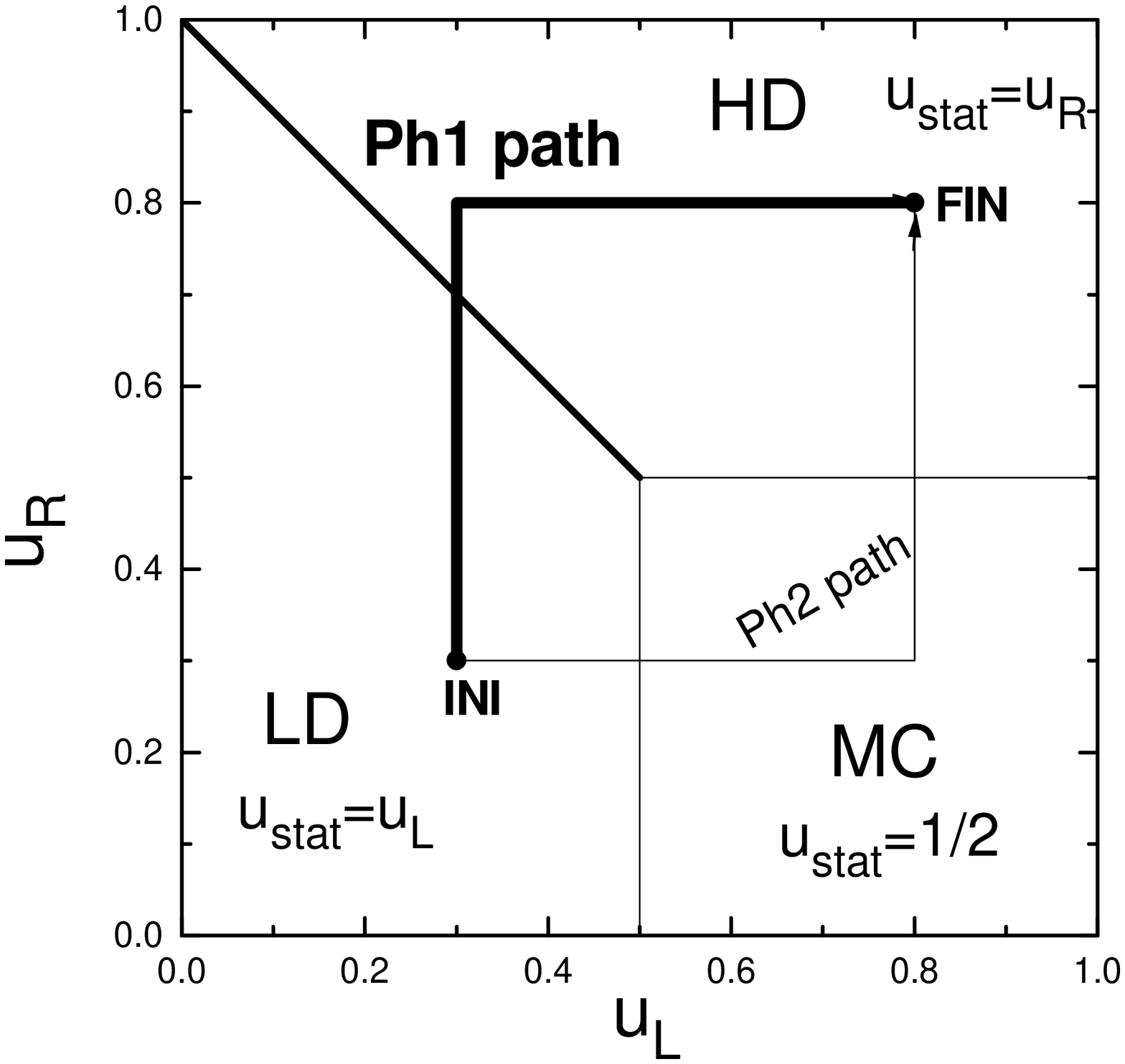}}
\subfigure[\label{ASEPPhase:b}]{
\includegraphics[width=4.3cm,height=4.5cm,clip]{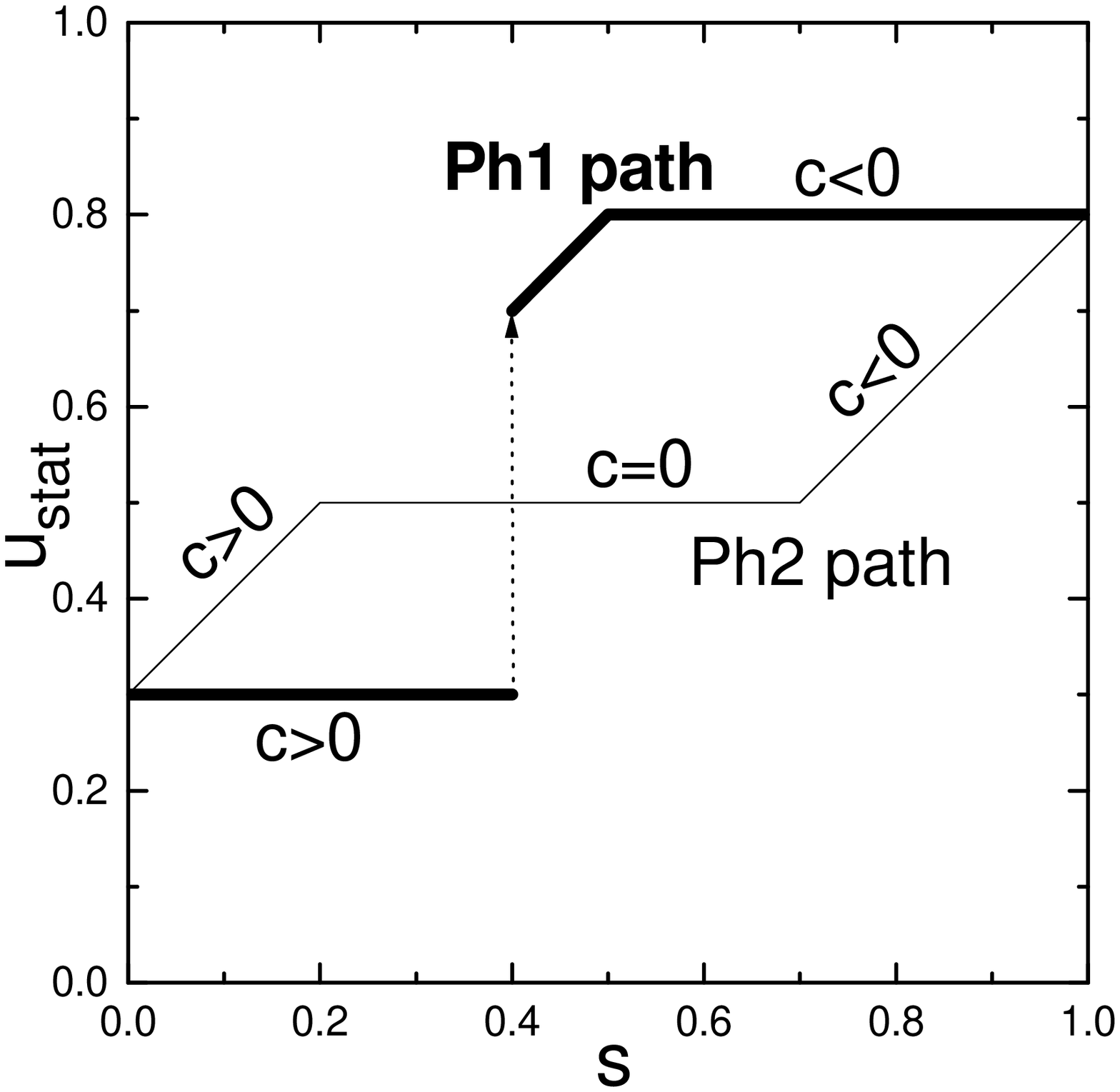}}
} \caption{\textbf{Panel (a)}: Phase diagram of TASEP with open boundaries.
LD, HD and MC denote Low Density, High Density and Maximal Current phases
respectively. Thick(thin) lines separating different phases mark phase
transitions of the first ( of the second) order. Thick and thin lines
originating in \textbf{INI} point (0.3,0.3) and ending in \textbf{FIN} points
show Ph1 and Ph2 paths, respectively. \textbf{Panel (b)}: Stationary densities
along the Ph1 path(bold line) and along the Ph2 path (thin line) shown on
Panel (a), versus running variable $s$, parametrizing the paths. A cusp in the
upper part of the Ph1 density path $u_{stat}(s)$ results from a cusp in the
Ph1 path in parameter space(Panel a).}%
\label{Fig_ASEPphase}%
\end{figure}

The path $\Gamma^{Ph1}(s)$ is continuous in the parameter space coordinates of
which are left and right boundary densities. We claim that at some point
$s^{\ast}$ whose precise location depends on the microscopic transition rates,
the stationary density $\rho_{stat}(s)$ undergoes discontinuous jump (a first
order transition) from some value $X\in G_{X}$ (non necessarily coinciding
with initial point $\rho^{ini}$) to a value $Y\in G_{Y}$ (non necessarily
coinciding with end point $\rho^{final}$). The stationary flux across the
transition is continuous, $j_{stat}(X)=j_{stat}(Y)$.

\textbf{Proposition II}: \textit{Two continuous (second order) phase
transitions of type }$p$\textit{ are observed along \textbf{any} path of type
Ph2. Those transitions are continuous phase transitions }$G_{X}\rightarrow
G_{X0Y}$\textit{ and }$G_{X0Y}\rightarrow G_{Y}$\textit{. } \begin{figure}[h]
\centerline{\scalebox{0.5}{\includegraphics{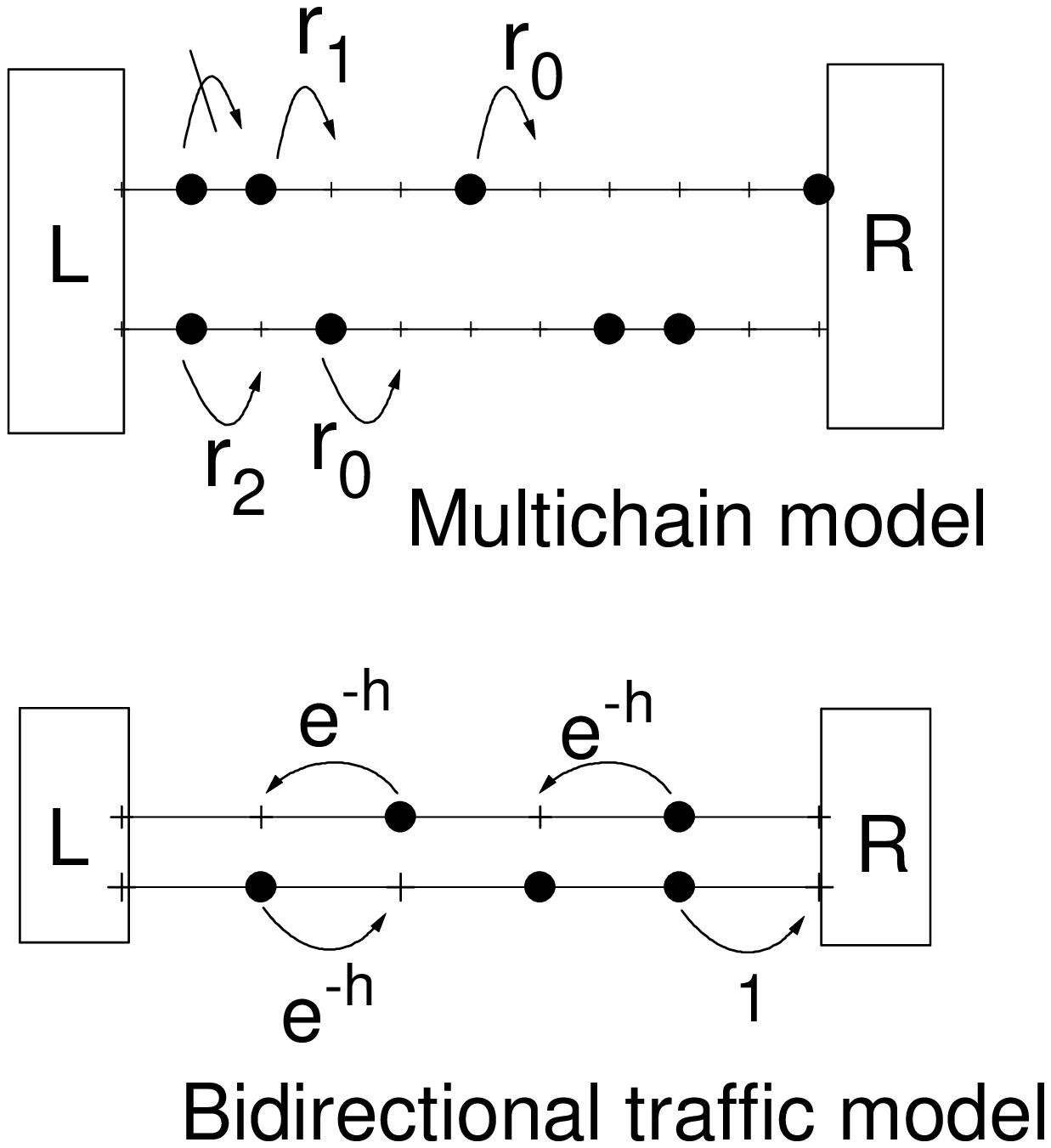}}
}\caption{Multi-chain model and two-way traffic model on a narrow road.
Coupling to boundary reservoirs is indicated by boxes marked L (left
reservoir) and R (right reservoir).}%
\label{Fig_TwoWayProcesses}%
\end{figure}

Summarizing Propositions I and II, we have that by proceeding along Ph1 (Ph2)
path connecting two stationary states in neighboring regions, we observe one
first order (two second order) transitions between these states (see
Fig.\ref{Fig_MinPath}). Alternatively, one can say that any Ph1 path in
parameter space maps onto a discontinuous path $\rho_{stat}(s)$ in physical
region, while any Ph2 path in parameter space maps onto a continuous path
$\rho_{stat}(s)$ in the physical region, a segment of which is pinned to the
$c_{p}=0$ hyperplane, see Fig.\ref{Fig_MinPath}. Continuous phase transitions
$G_{X}\rightarrow G_{X0Y}$ and $G_{X0Y}\rightarrow G_{Y}$ correspond to
pinning and depinning points of the Ph2 trajectory. We remark that in particle
systems without hysteresis the above paths are fully invertible, i.e. by
proceeding along a path $\Gamma(s)$ in the opposite direction from the final
to initial point, one encounters exactly the same mapping $\Gamma
(s)\rightarrow\rho_{stat}(s)$.

The physical significance of the Ph2 path resides in the fact that it contains
favourable boundary setups for the formation of rarefaction waves governing
the $G_{X0Y}$-type stationary states (Ph2 segment marked in
Fig.\ref{Fig_MinPath}(a) by dashed line) and does not contain boundary setups
favoring stable shock waves which govern discontinuous phase transitions.
Consequently, discontinuous shock-wave driven transitions are impossible along
any Ph2 path and the mapping $\Gamma(s)\overset{Ph2}{\rightarrow}\rho
_{stat}(s)$ is continuous (see also
sec.\ref{sec::Rarefaction waves govern continuous phase transitions}).
Analogously, one can deduce that the mapping $\Gamma(s)\overset{Ph1}%
{\rightarrow}\rho_{stat}(s)$ along any Ph1 path must be discontinuous. More
details are given in
Secs.\ref{sec::Shock waves interaction as a mechanism of the phase transitions of the first order}%
,\ref{sec::Rarefaction waves govern continuous phase transitions}. Before
discussing the kinetic mechanisms underlying the transitions we illustrate
Propositions I and II with some specific examples.

\textbf{Case K=1.} It is instructive to start with the simplest possible and
well known case of one particle species $K=1$. Let us take the Totally
Asymmetric Simple Exclusion Process, or TASEP \cite{Schu00}%
,\cite{TASEP_reviewDerrida}  as a representative. This process is defined on a
chain, each site of which can be empty or occupied by one particle. Particles
jump independently after an exponentially distributed random time with mean
$1$ to a nearest neighbor site on the right, provided that the target site is
empty (hard core exclusion rule). We recall that the particle flux, which is a
number of particles crossing a single bond per unit time,  as function of the
density $u$ for this model has the form $j(u)=u(1-u)$ and the physical region
of densities is $0\leq u\leq1$ due to the hard core exclusion.
\begin{figure}[h]
\centerline{
\includegraphics[width=5.5cm,height=5.5cm,clip]{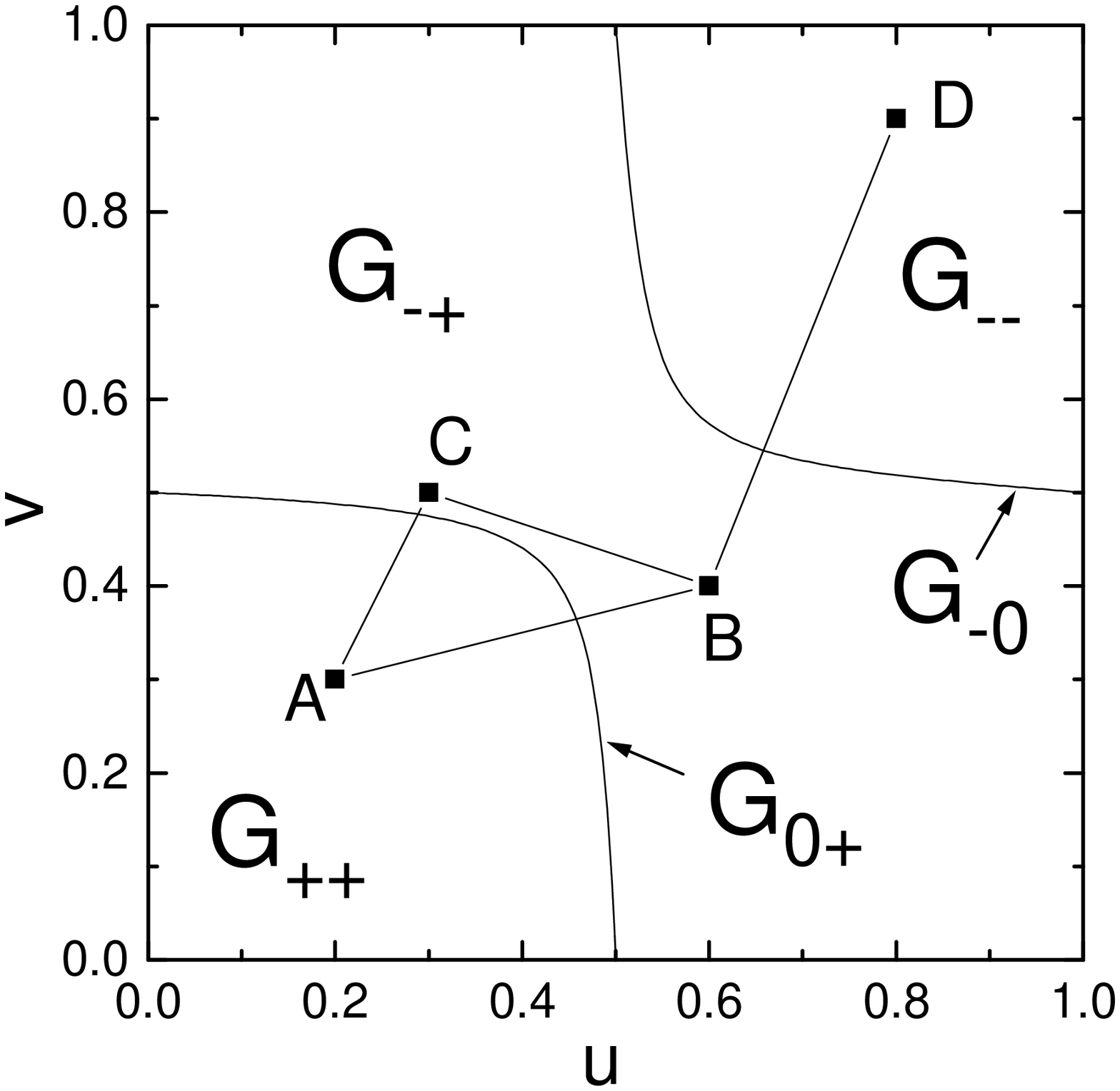}
}\caption{Decomposition of the physical domain for a model with two species [
multi-chain model with $K=2,\gamma=0.5$, see text for model definition],
according to signs of characteristic speeds $c_{1},c_{2}$. E.g. $G_{0+}$
denotes region where $c_{1}=0,\ c_{2}>0$ etc.. Points $A,B,C,D$ are reference
points in respective $G$-domains, chosen for illustration of Ph1 and Ph2 paths
in Figs.\ref{Fig_Ph},\ref{Fig_PhPh}.}%
\label{Fig_Gdecomposition}%
\end{figure}The characteristic speed $c=j^{\prime}(u)=1-2u$ is positive for
$u\in\lbrack0,1/2)$, is negative for $u\in(1/2,1]$, and it vanishes  for
$u=1/2$, thus defining the respective domains $G_{+},G_{-}$ and $G_{0}$. The
boundary densities are $u_{L}$ and $u_{R}$. The well-known stationary states
of TASEP with open boundaries, the Low Density (LD) state, the High Density
(HD) state, and the Maximal Current (MC) state are readily identified with
$G_{+}$, $G_{-}$ and $G_{0}$ states, respectively. The phase diagram of TASEP
and the mappings $\Gamma(s)\rightarrow u_{stat}(s)$ for Ph1- and Ph2-paths are
shown in Fig.\ref{Fig_ASEPphase}. Note that along the Ph1\ path one encounters
a discontinuous phase transition from LD to HD state $G_{+}\rightarrow G_{-}$.
Along the Ph2 path two continuous phase transitions $G_{+}\rightarrow G_{0}$ ,
$G_{0}\rightarrow G_{-}$ take place.

\textbf{Case K=2.} As examples we choose a multi-chain model
\cite{Mario_multiASEP} restricted to $K=2$, and a two-way traffic model on a
narrow road. The former describes interacting exclusion processes evolving on
parallel chains with species hopping in the same direction, while the latter
describes the exclusion process with species hopping in opposite direction,
see Fig.\ref{Fig_TwoWayProcesses}. Both models are described in continuous
limit by equations of type (\ref{PDE}),(\ref{PDEBoundaryConditions}).

The multi-chain model is solvable on a ring, its stationary
current is known analytically together with the diffusion matrix
$B$, this giving the possibility of using either hydrodynamic
limit equations (\ref{PDE}) or microscopic approach (Monte Carlo
simulations). Moreover, it is known from previous studies
\cite{reflections_JPA}, \cite{Mario_multiASEP}, that numerical
integrations of the respective discretized hydrodynamic equations
(\ref{PDE}), (\ref{PDEBoundaryConditions}), reproduce very
accurately (if not exactly) the outcome of the respective
Monte-Carlo simulations. The other model (two way traffic) is not
solvable and Monte Carlo simulations remain the only tool of
investigation. \begin{figure}[ptbh] \centerline{
\subfigure[\label{figPh:a}]
{\includegraphics[width=4.25cm,height=4.25cm,clip]{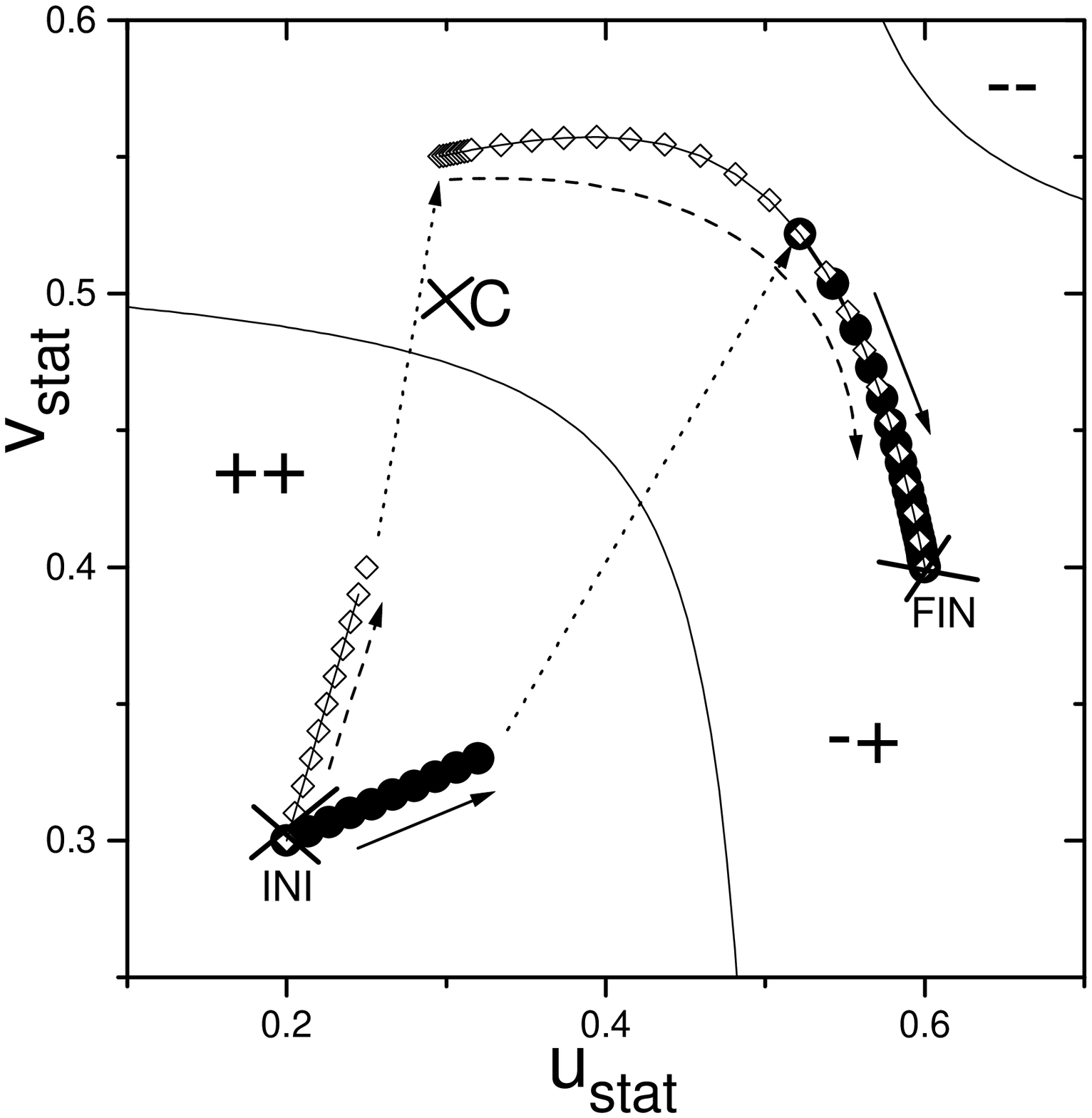}}
\qquad \subfigure[\label{figPh:b}]
{\includegraphics[width=4.25cm,height=4.25cm,clip]{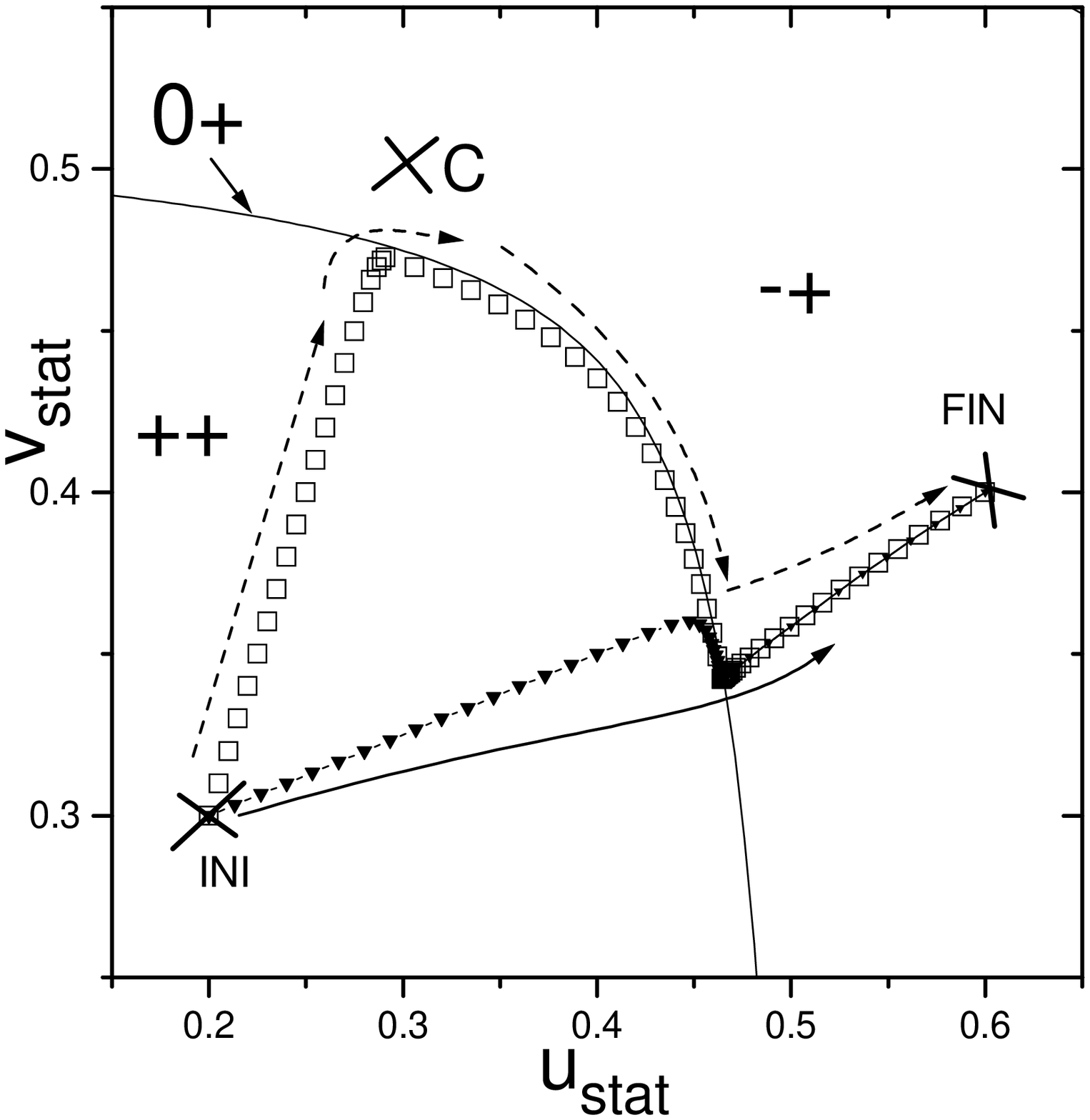}}
\qquad } \caption{Location of stationary densities for two-chain
model ($\gamma=0.5$) for (a) two alternative Ph1 paths (b) two
alternative Ph2 paths from $G_{++}$ to $G_{-+}$ domains. Initial
and end points for all paths corresponds to fully matching
left-right boundary densities $\rho _{ini}=(0.2,0.3)$,
$\rho_{final}=(0.6,0.4)$ (points A,B in
Fig.\ref{Fig_Gdecomposition}). Path I (filled symbols) is a direct
path $\rho_{ini}\rightarrow\rho_{final}$, where the boundary
densities during respective steps L and R (see Sec.\ref{sec::A
minimal path}) change by linear interpolation, e.g.
$u_{R}(s)-u_{R}^{ini}=(u_{R}^{fin}-u_{R}^{ini})s$, etc.. Path II
(open symbols) goes through intermediate point C with coordinates
$\rho_{C}=(0.3,0.5)$ i.e. each step L,R involves sequence of two
linear
interpolations $\rho_{ini}\rightarrow\rho_{C},\rho_{C}\rightarrow\rho_{final}%
$. Evolution of the densities along the paths I and II is indicated by solid
and dashed arrows, respectively, crosses mark initial, intermediate and final
points. Data points are taken from numerical integration of the respective
discretized hydrodynamic equations (\ref{PDE}),(\ref{PDEBoundaryConditions}).
Deviations from the exact $c_{p}=0$ line in Panel (b) are due to finite-size
errors.}%
\label{Fig_Ph}%
\end{figure}

First consider the multi-lane model for $K=2,$
%, which describes two coupled
%exclusion processes on parallel chains
see the top panel of Fig.\ref{Fig_TwoWayProcesses}. The only allowed move is a
hopping of a particle to its nearest neighbouring site on its right, with the
rate $r_{n}=1-n\gamma/2$ where $0\leq n\leq2$ is the number of particles on
the adjacent chain, neigbouring to the departure and to the target site (see
Fig.\ref{Fig_TwoWayProcesses}). The interchain interaction parameter $\gamma$
varies from $\gamma=0$ (corresponding to two uncoupled TASEPs ) to $\gamma=1$
(maximal interaction). Particle hopping obeys the exclusion rule: if the
target site is already occupied, the move is rejected. The hopping between
chains is not allowed. We denote the average density of particles on the first
and second chain as $u,v$, respectively. Due to hardcore exclusion, the
physical region of densities for this model is the square domain $0\leq
u,v\leq1$. The model has product stationary states, which allows to calculate
stationary currents \cite{Mario_multiASEP} $j_{u}=u(1-u)(1-\gamma v)$ and
$j_{v}=v(1-v)(1-\gamma u).$ The characteristic velocities $c_{k}$ are the
eigenvalues of the flux Jacobian $(D\mathbf{j})\varphi_{k}=c_{k}\varphi_{k}$,
where%
\begin{equation}
D\mathbf{j}=%
\begin{pmatrix}
(1-2u)(1-\gamma v) & -\gamma u(1-u)\\
-\gamma v(1-v) & (1-2v)(1-\gamma u)
\end{pmatrix}
. \label{Jacobian}%
\end{equation}
Consistently with our notations, a domain in $u,v$ plane with $c_{1}(u,v)>0,$
$c_{2}(u,v)>0$ will be denoted by $G_{++}$ (and similarly for other sign
combinations). For generic interaction $\gamma<1$, all domains $G_{++}%
,G_{-+},G_{--}$ are present, see Fig.\ref{Fig_Gdecomposition}. As $\gamma$
increases, the domain $G_{--}$ shrinks and for $\gamma=1$ (maximal
interaction) it disappears completely.

According to the Proposition I, described in the Sec.\ref{sec::A minimal path}%
, starting from the domain $G_{++}$ and going to the domain $G_{-+}$ using a
Ph1 path (\ref{Ph1}), we should observe discontinuous transition in stationary
density along the path. In Fig. \ref{figPh:a} stationary densities along two
alternative Ph1 paths between the same initial point (located in $G_{++}$) and
final point (located in $G_{-+}$) are shown. These paths, indicated as $AB$
and $ACB$ in Fig.\ref{Fig_Gdecomposition}, differ by the elementary
trajectories $\rho_{ini}\rightarrow\rho_{final}$ are built, see caption of
Fig.\ref{figPh:a}. As expected, the qualitative scenario does not depend on
the details of the path: in both cases we see a discontinuous transition
$G_{++}\rightarrow G_{-+}$. Similarly, along the two alternative Ph2 paths
(\ref{Ph2}) two continuous BDPT occur: $G_{++}\rightarrow G_{0+}$ and
$G_{0+}\rightarrow G_{-+}$ (see Fig.\ref{figPh:b}), in accordance with the
Proposition 2. \begin{figure}[ptb]
\centerline{
\subfigure[\label{fig:a}]{\includegraphics[width=4.25cm,height=4.25cm,clip]{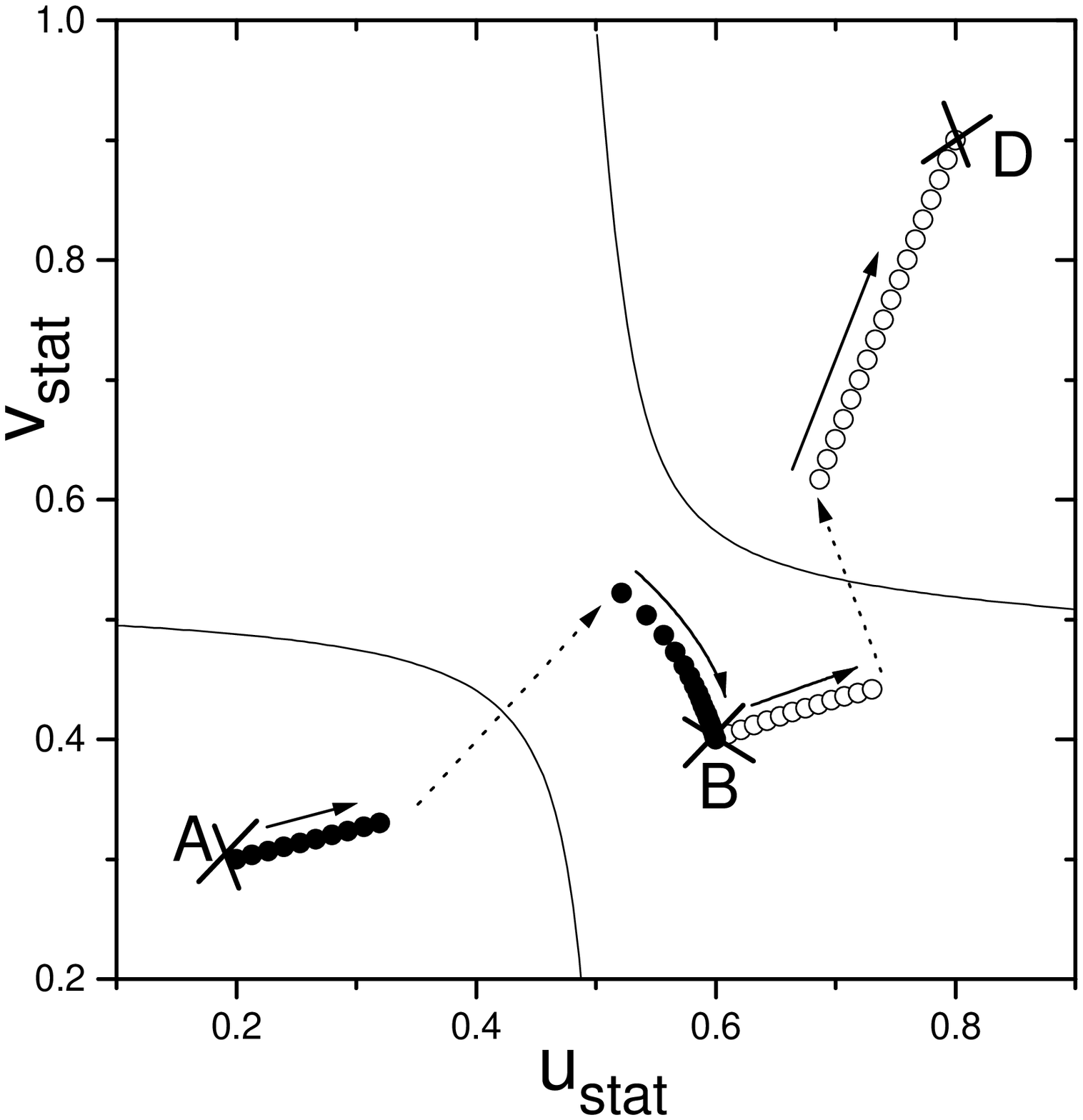}}
\subfigure[\label{fig:b}]{\includegraphics[width=4.25cm,height=4.25cm,clip]{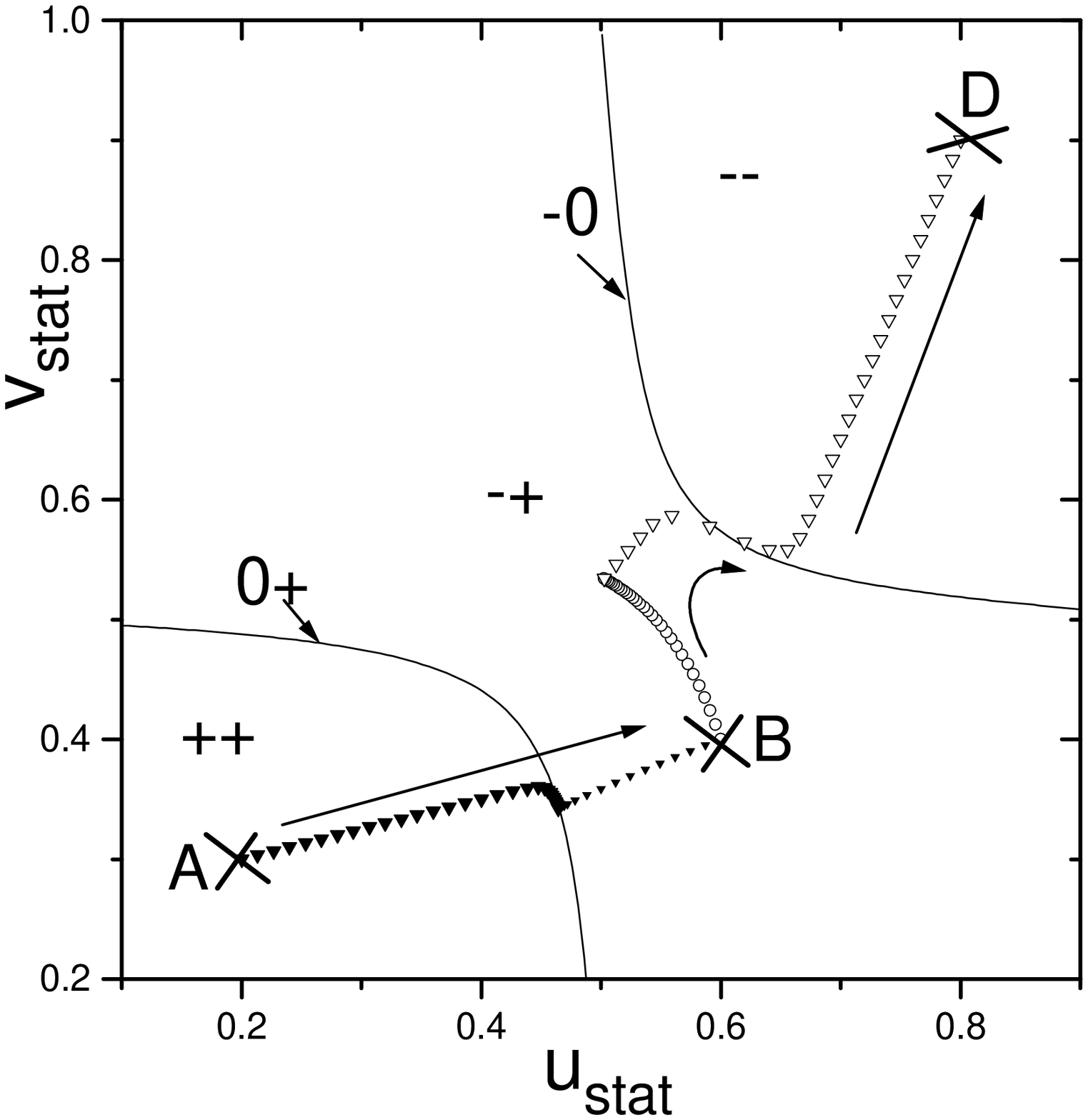}}
} \centerline{
\subfigure[\label{fig:c}]{\includegraphics[width=4.25cm,height=4.25cm,clip]{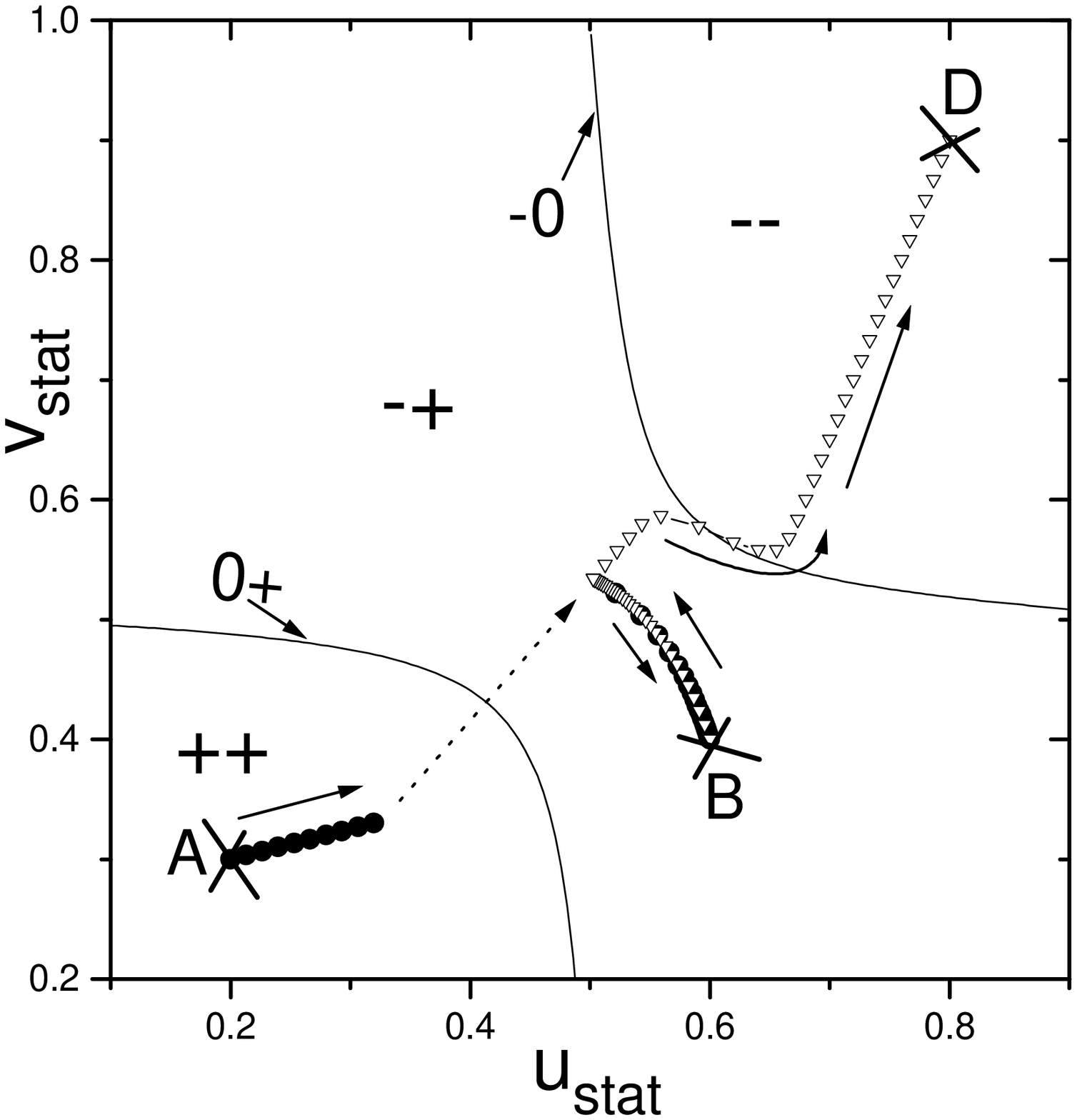}}
\subfigure[\label{fig:d}]{\includegraphics[width=4.25cm,height=4.25cm,clip]{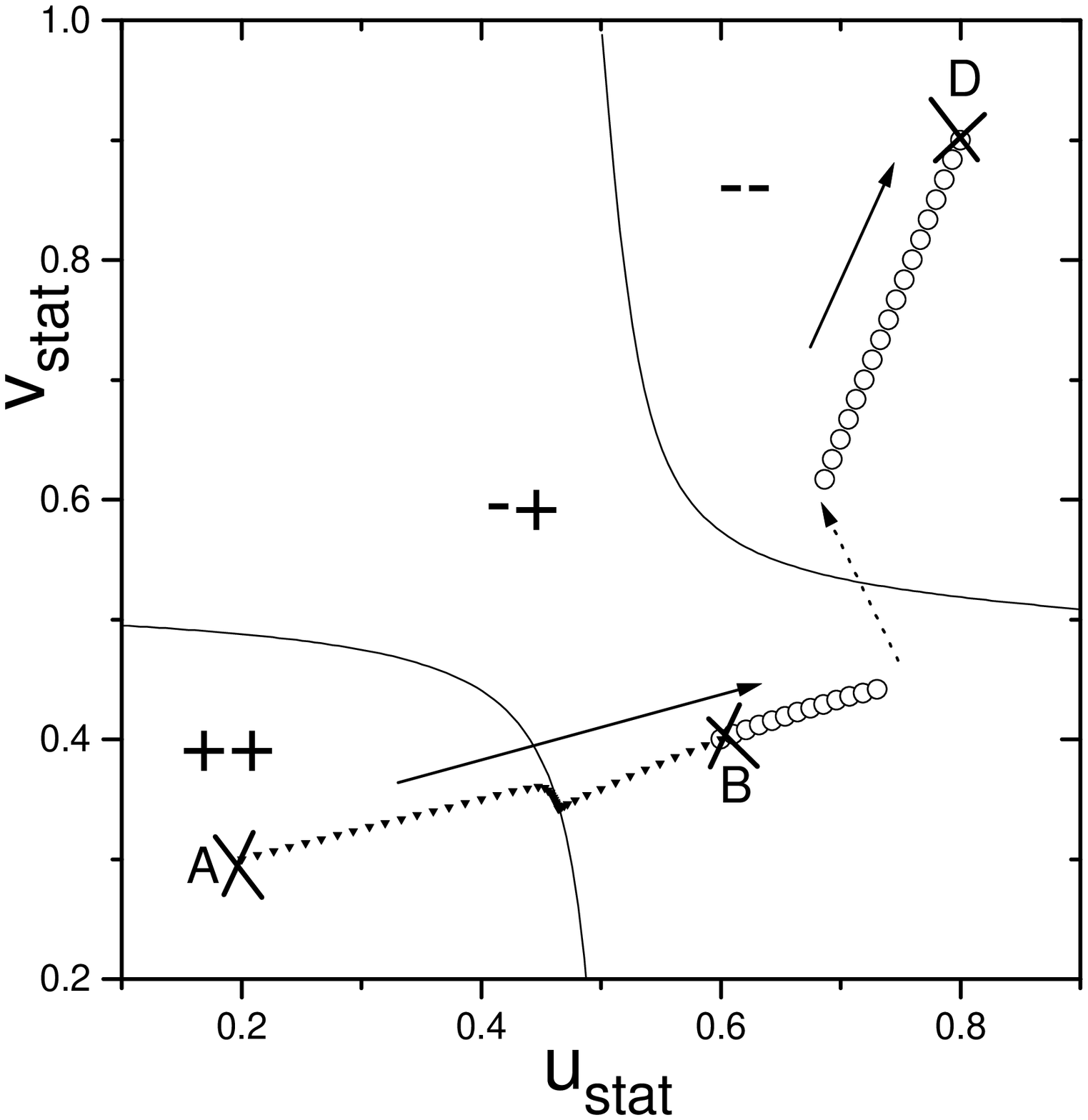}}
}\caption{Panels (a)-(d): Location of stationary densities for two-chain model
with $\gamma=0.5$ along (Ph1)(Ph1),(Ph2)(Ph2),(Ph1)(Ph2), and (Ph2)(Ph1)
paths, respectively. All trajectories pass through initial (A), middle (B) and
final (D) points. E.g. Panel (b) shows stationary densities along two
consecutive Ph paths $A\rightarrow^{Ph1}B$,$B\rightarrow^{Ph2}D$. Filled
(empty) symbols mark stationary densities $u_{stat},v_{stat}$ along the first
(the second) Ph path. Evolution of the densities along the path is shown by
arrows. Along every Ph2 piece a pinning to/depinning from $c_{p}=0$ line is
observed (two continuous phase transitions). Along every Ph1 piece a
discontinuous phase transition occurs, marked by dotted arrows. }%
\label{Fig_PhPh}%
\end{figure}

\subsection{Building composite paths}

From the point B in $G_{-+}$ where all above described minimal paths ended, we
can continue further building Ph1 $G_{-+}\overset{Ph1}{\rightarrow}G_{--}$
(\ref{Ph1}) or Ph2 $G_{-+}\overset{Ph2}{\rightarrow}G_{--}$ (\ref{Ph2}) paths
to some point D in $G_{--}$, see Fig.\ref{Fig_Gdecomposition}.
\begin{figure}[h]
\centerline{\scalebox{0.4}{\includegraphics{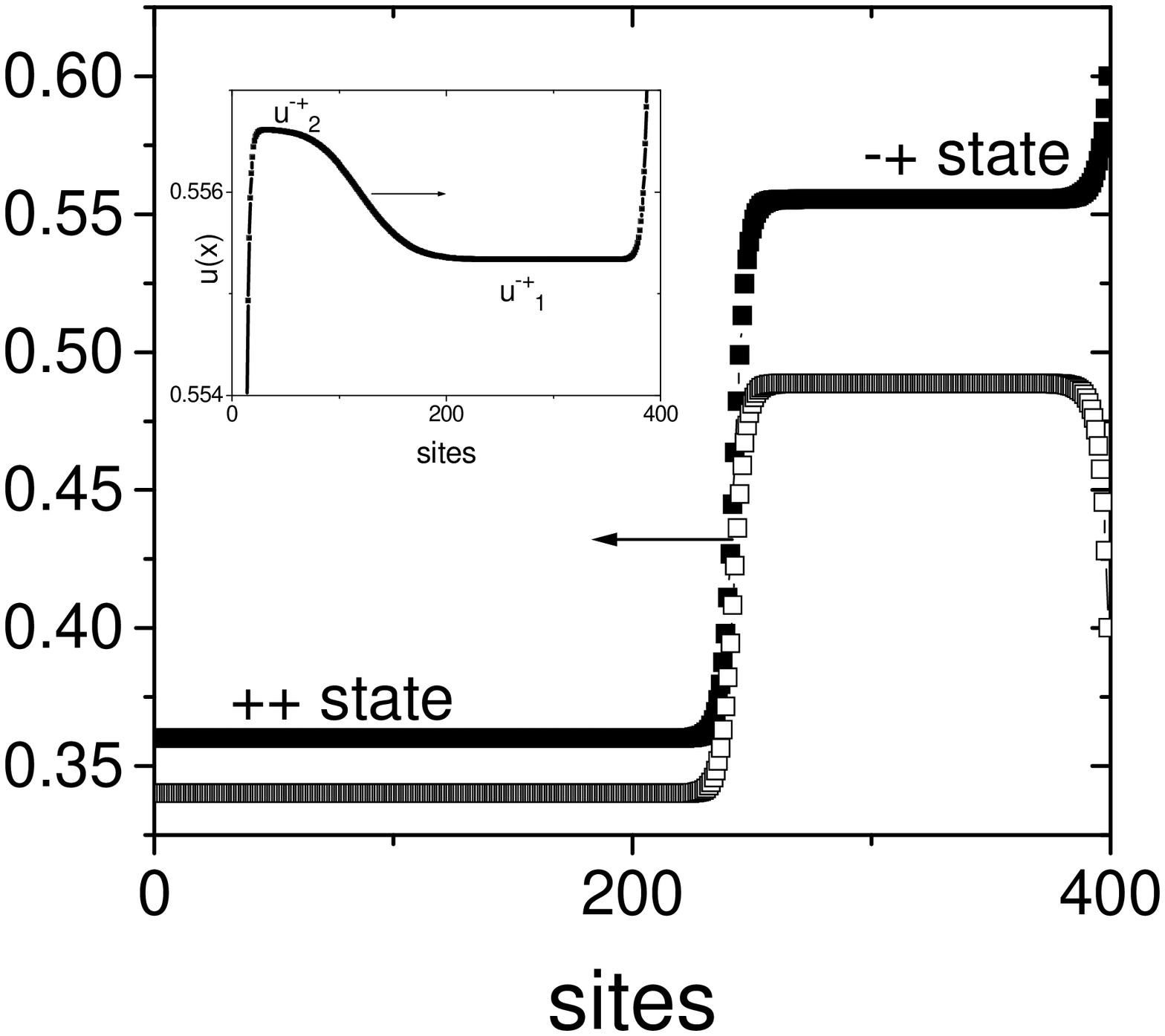}}
}\caption{Average density profiles $u(x,t),v(x,t)$ evolution close to the
first order phase transition line. Initial state at $t=0$ is a state matching
the left boundary $u_{L}=0.36,v_{L}=0.34$(the right boundary is $u_{R}%
=0.6,v_{R}=0.4$). As time goes on, a shock wave at the right boundary appears
and starts to propagate, its position at $t=6400$ is shown. After reaching the
left boundary, it reflects, see Inset, taken at $t=17600$. Final stationary
state has densities $u=0.5568,v=0.4868$.}%
\label{Fig_Ph1Mechanism}%
\end{figure}Along the new Ph1 (Ph2) path from B to D we shall see another
discontinuous (continuous) transition in the stationary density. Combining all
possible Ph1 and Ph2 paths leading from $G_{++}$ $\rightarrow G_{-+}%
\rightarrow$ $G_{--}$, one can observe a desired sequence of phase
transitions. E.g. choosing $\ $only Ph1 paths $G_{++}\overset{Ph1}%
{\rightarrow}G_{-+}\overset{Ph1}{\rightarrow}G_{--}$, we see two phase
transitions of the first order, see Fig.\ref{fig:a}. Along a path
$G_{++}\overset{Ph2}{\rightarrow}G_{-+}\overset{Ph1}{\rightarrow}G_{--}$ we
see two continuous transitions $G_{++}\rightarrow G_{0+}\rightarrow G_{-+}$
and a discontinuous transition $G_{-+}\rightarrow G_{--}$, etc.. The outcome
of all four possible choices of composite paths passing from $G_{++}$ to
$G_{--}$ are shown in Figs.\ref{fig:a}-\ref{fig:d}.

\subsection{BDPTs for arbitrary number of species}

The generalization of these considerations to systems with arbitrary number of
species is straightforward. In systems with $K$ particle species, one can
construct composite paths leading from a state with all positive
characteristic speeds $G_{+...+}$ to a state with all negative characteristic
speeds $G_{-...-}$, similarly to those shown in Figs.\ref{fig:a}-\ref{fig:d}.
Every composite path consists of $K$ consecutive minimal Ph1 or Ph2 paths,
described in Sec.\ref{sec::A minimal path}. Obviously, the number of
qualitatively different composite paths is $2^{K}$. Along any composite path
from $G_{+...+}$ state to $G_{-...-}$ state we shall observe at least $Z$
first order transitions, and at least $2(K-Z)$ discontinuous phase
transitions, where $Z$ is the number of \ Ph1 pieces in the composite path. We
say "at least", because e.g. the number of first order transitions can
occasionally be larger than $Z$ (in case of complicated shape of $G$-domains
or if composite path crosses the same hyperplane $c_{p}=0$ several times), but
not smaller than $Z$, and analogously for second order transitions.
\begin{figure}[h]
\centerline{\scalebox{0.4}{\includegraphics{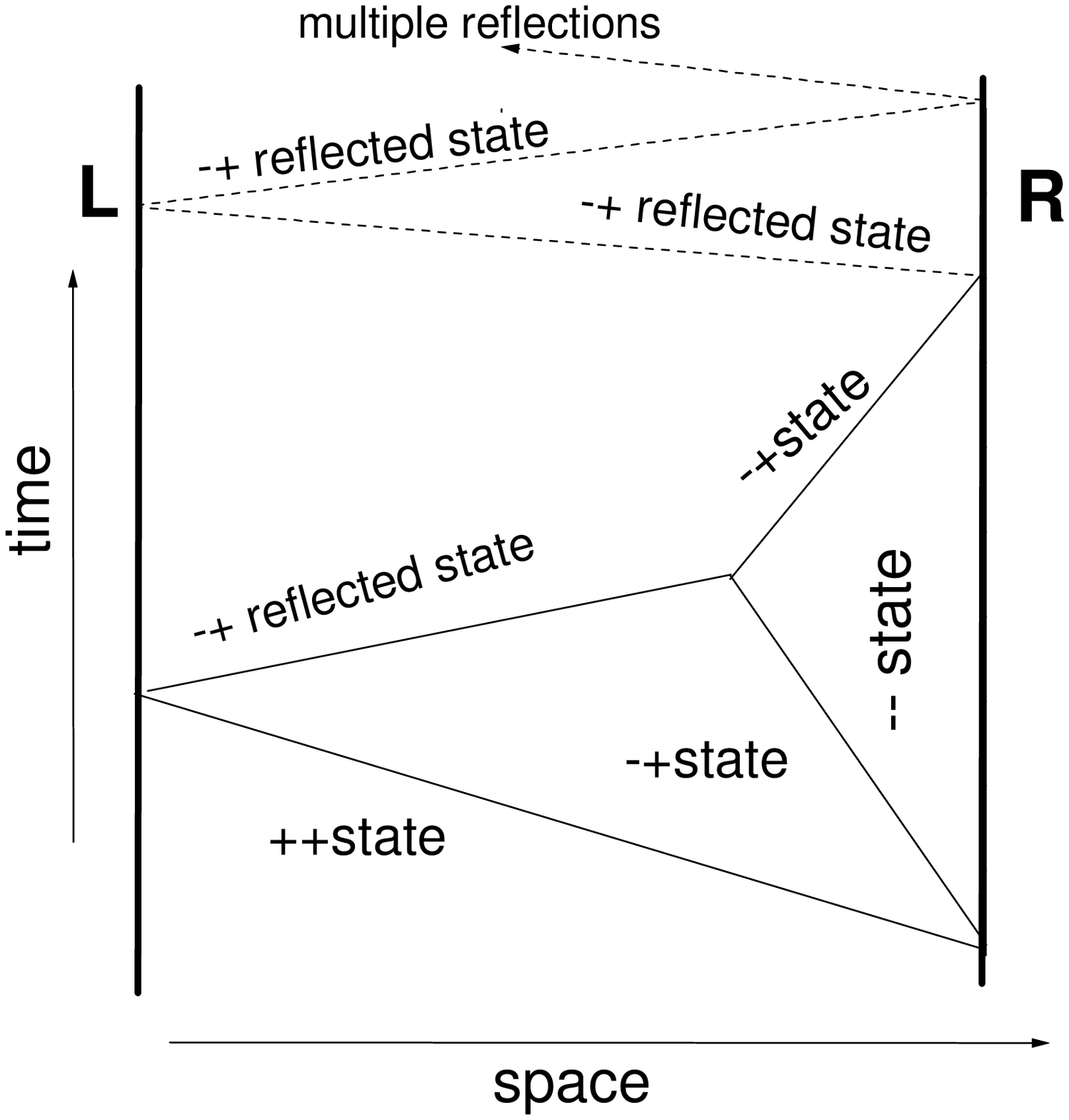}}
}\caption{Schematic picture of a shock waves interaction scenario for
two-species system. Dashed line denote multiple reflections of the $-+$ shock
from the boundaries, during which the bulk density exponentially converges to
its stationary value. }%
\label{Fig_Ph1MechComplex}%
\end{figure}E.g. already in one-component systems with a double maximum in the
current-density relation $j(u)$, both $G_{+}$ and $G_{-}$ domains consist of
two separated segments and a Ph1 path between disjoint $G_{+}$ and $G_{-}$
domains may lead to observation of two discontinuous transitions. In the next
two sections we discuss the kinetic mechanisms governing BDPTs along minimal
Ph1 and Ph2 paths.

\section{Shock wave mechanism underlying first order BDPT}

\label{sec::Shock waves interaction as a mechanism of the phase transitions of the first order}%

Consider the first order phase transitions shown in Fig.\ref{figPh:a} from the
$G_{++}$ to $G_{-+}$ state. Initial $G_{++}$ state is a state perfectly
matching the left boundary. Note that the perfect match with the left boundary
is the only possible way for a $G_{++}$ state to be stationary because any
eventual perturbation at the left boundary will be carried away from it since
both characteristic velocities are positive (see also \cite{PopkovCambridge}).
As we cross a transition point (at a small but finite distance from it) a new
$-+$ shock wave with densities $u_{1}^{-+},v_{1}^{-+}$ belonging to $G_{-+}$
appears at the right boundary and starts to propagate inside the bulk. Note
that the new shock does not match the right boundary $u_{1}^{-+}\neq
u_{R},v_{1}^{-+}\neq v_{R}$) but forms a boundary layer with it (see
Fig.\ref{Fig_Ph1Mechanism}). The densities $u_{1}^{-+},v_{1}^{-+}$ are not the
final stationary densities yet. After hitting the left boundary the shock
reflects and changes its density to another value, $u_{2}^{-+},v_{2}^{-+}$ ,
see Fig. \ref{Fig_Ph1Mechanism}. This reflected wave, in its turn, hits the
right boundary, reflects again and changes its density to $u_{3}^{-+}%
,v_{3}^{-+}$.

This process continues indefinitely and the sequence $\{u_{n}^{-+},v_{n}%
^{-+}\}$ converges exponentially to the stationary value $u_{stat}%
^{-+},v_{stat}^{-+}$ as $n\rightarrow\infty$. The shock densities $u_{n}%
^{-+},v_{n}^{-+}$ for odd $n=1,3,...$ (for even $n=2,4,...$) belong to the
right reflection map defined by $u_{R},v_{R}$ (to the left reflection map
defined by $u_{L},v_{L}$), see \cite{reflections_JPA}, \cite{reflections_ABO}
for more details. In practice, it becomes harder and harder to observe
reflections of high order since the respective shocks differ by an
infinitesimal change of densities (see inset of Fig.\ref{Fig_Ph1Mechanism}).
The sequence $\{u_{n}^{-+},v_{n}^{-+}\}$ , (unlike the fact of a presence of
first order transition along the Ph1 path) is microscopic rates dependent,
through the diffusion matrix $B$ from (\ref{PDE}).

It is important to stress that precisely at the transition point the shock
wave between the $++$ state (matching the left boundary $u_{L},v_{L}$) and the
$-+$ state with densities $u_{1}^{-+},v_{1}^{-+}$ is unbiased, meaning that
there is a perfect balance between the respective currents: $j_{u}(u_{L}%
,v_{L})=j_{u}(u_{1}^{-+},v_{1}^{-+}),$ $j_{v}(u_{L},v_{L})=j_{v}(u_{1}%
^{-+},v_{1}^{-+})$. The fact that $c_{1}(u_{L},v_{L})>0$ ($c_{1}(u_{1}%
^{-+},v_{1}^{-+})<0$) at the left (right) from shock discontinuity guarantees
the stability of the unbiased shock $G_{++}/G_{-+}$. At one side of the phase
transition this shock is biased to the right leading to $G_{++}$ stationary
state, and at another side of the phase transition it is biased to the left
leading to $G_{-+}$ stationary state. This feature is essentially the same as
in one-species systems, see \cite{Kolo98}. Precisely at the phase transition
line the unbiased shock performs a random walk between the boundaries. By
averaging the local particle density over large times one samples
configurations with the shock at all possible positions, which leads to a
density profile with a linear slope, observed in Monte Carlo simulations (not
shown for brevity).

More complicated scenarios at first order transition points may be observed if
we choose left and right boundary densities belonging to non-connected $G$
domains (e.g. the left boundary belongs to $G_{++}$ and the right boundary
belongs to $G_{--}$ domain), and impose flat initial conditions matching one
of the boundaries, e.g. the left boundary of $G_{++}$ type. Close to the phase
transition $G_{-+}$ $\rightarrow$ $G_{--}$ shown at Fig.\ref{Fig_IniFinalAll}
along a single big Ph1 path, we shall see appearance of two shocks at the
right boundary of $++/-+$ and of $-+/--$ type. For a while in the system there
are two moving consecutive shocks.

Stability of such a \ multi-shock is possible due to existence of two
conserved quantities, see \cite{Lax73}. The first shock reaches the left
boundary and reflects from it (with changed densities). Now we have two shocks
which counter-propagate and collide at some point, forming a single shock of
$-+/--$ type. The future stationary state is then determined by the direction
of motion of this shock: for positive (negative) shock velocity the resulting
stationary state will be of $-+$ (of $--$ type). Again, we see that the first
order phase transition is caused by the direction of the bias of a shock
connecting two states. The schematic space-time evolution of the above
described scenario is shown in Fig.\ref{Fig_Ph1MechComplex}. Note, however,
that if we follow an adiabatic path, the initial conditions as we have imposed
will not appear (close to a transition point $G_{-+}$ $\rightarrow$ $G_{--}$
the initial state, to be quasi-stationary, must be either of $G_{-+}$ or of
$G_{--}$ type), and consequently at any time we shall see at most one shock in
the system.

This scenario of the first order phase transitions described above for two
particle species is straightforwardly generalizable to an arbitrary number of
species $K$. A discontinuous phase transition of the type $p$ (see
Sect.\ref{sec::Hierarchy of continuous and discontinuous phase transitions})
along the minimal path described in Sec.\ref{sec::A minimal path} is caused by
a shock $G_{X}/G_{Y}$ between a $G_{X}$ state on the left from discontinuity
and a $G_{Y}$ state on the right from discontinuity. State $G_{X}$ has one
extra positive characteristic speed ($c_{p}>0$) with respect to the $G_{Y}$
state ($c_{p}<0$), see (\ref{G domains}).

%\textbf{Table 1}. Stationary bulk densities along a minimal Ph1 path from
%$G_{++}$ to $G_{-+}$
\begin{table}[ptb]
$_{%
\begin{tabular}
[c]{|l|l|l|l|l|l|}\hline
Left boundary & $++$ $\rightarrow$ & $++$ $\rightarrow$ & $++$ $\rightarrow$ &
$0+$ $\rightarrow$ & $-+$\\\hline
Right boundary & $++$ $\rightarrow$ & $0+$ $\rightarrow$ & $-+$ $\rightarrow$
& $-+$ $\rightarrow$ & $-+$\\\hline
Stationary bulk & $++$ & $++$ & $++/-+$ & $-+$ & $-+$\\\hline
\end{tabular}
}$\caption{Sequential changes of the stationary state densities along a
minimal path Ph1 (\ref{Ph1}) from $G_{++}$ to $G_{-+}$.}%
\end{table}

The signs of all remaining characteristic speeds (but not the characteristic
speeds themselves) are the same at both sides of the shock. At the transition
point, the shock $G_{X}/G_{Y}$ is unbiased (has zero velocity) and its
stability is guaranteed by the fact that $c_{p}>0$ ($c_{p}<0$) on the left (on
the right) from the discontinuity. Such a shock is called a $p$-shock in the
PDE theory of conservation laws \cite{Lax73}. Zero shock velocity signalizes
equality of particle currents of all species at both sides of discontinuity.
Consequently, the stationary current is continuous across the transition.

At one side of the transition, the $G_{X}/G_{Y}$ shock is biased to the right
and then the $G_{X}$-type stationary state prevails. At the other side of the
transition the shock is biased to the left, resulting in the $G_{Y}$-type
stationary state. Thus, the density changes discontinuously at the transition
point while the current is continuous (usually it has a cusp) across the
transition point. The location of the transition point itself can be indicated
on the respective Ph1 path only approximately, as being inside the dashed-like
decorated segment of it, see Fig.\ref{Fig_MinPath}(a)) and Table 1.

If the left and right boundary densities belong to non-connected $G$ domains
(but the number of positive characteristic speeds at the left boundary is
larger than those at the right boundary) and if the initial state of the
system matches one of the boundaries, a stable multiple shock can be observed.
The stationary state of the system is then decided by a sequence of
reflections from the boundaries (governed by the diffusion matrix $B$ in
(\ref{PDE})), and interactions in the bulk between shocks (governed by the
current-density relations $j_{q}(u_{1},u_{2},...,u_{K})$).

\section{Rarefaction wave mechanism underlying second order BDPT}

\label{sec::Rarefaction waves govern continuous phase transitions}

If a $G_{X}/G_{Y}$ shock is stable (see preceeding Section), the inverse shock
$G_{Y}/G_{X}$ is unstable and gives rise to a rarefaction wave which is a
self-similar solution of (\ref{PDE}), depending only on ratio $\xi
=(x-x_{0})/t$ where $x_{0}$ is a position of its center, and $t>0$. Let us
argue that in the long-time limit $t\rightarrow\infty$ the stationary bulk
density $\mathbf{u}_{stat}$ generated by a rarefaction wave, has zero
characteristic speed $c_{p}(\mathbf{u}_{stat})=0$. By $\mathbf{u}_{stat}$ we
denote a set of bulk stationary densities $\{u_{1}^{stat},u_{2}^{stat}%
,...,u_{K}^{stat}\}$. We search for a solution of (\ref{PDE}) in the form
$u(x,t)=h(\xi)$. Substituting in (\ref{PDE}), and denoting the derivative with
respect to $\xi$ with a prime, we obtain
\begin{equation}
-\frac{\xi}{t}h^{\prime}+\frac{1}{t}(D\mathbf{j})h^{\prime}=\frac{1}{t^{2}%
}O(\varepsilon),
\end{equation}
where the matrix $(D\mathbf{j})(h(\xi))$ is the Jacobian of the flux
$(D\mathbf{j})_{pq}=\partial j_{p}/\partial u_{q}$. The above equation can be
rewritten as
\begin{equation}
(D\mathbf{j})h^{\prime}=\xi h^{\prime}+\frac{O(\varepsilon)}{t}.
\end{equation}
\begin{table}[ptb]
$%
\begin{tabular}
[c]{|l|l|l|l|l|l|}\hline
Left boundary & $++$ $\rightarrow$ & $0+$ $\rightarrow$ & $-+$ $\rightarrow$ &
$-+$ $\rightarrow$ & $-+$\\\hline
Right boundary & $++$ $\rightarrow$ & $++$ $\rightarrow$ & $++$ $\rightarrow$
& $0+$ $\rightarrow$ & $-+$\\\hline
Stationary Bulk & $++$ & $0+$ & $0+$ & $0+$ & $-+$\\\hline
\end{tabular}
$\caption{Sequential changes of the stationary state densities along a minimal
path Ph2 from $G_{++}$ to $G_{-+}$.}%
\end{table}

In the limit $t\rightarrow\infty$ the $O(\varepsilon)/t$ term vanishes,
$\xi=(x-x_{0})/t\rightarrow0$ for any finite $x$, and the above equation
reduces to $\left.  (Dj)\right\vert _{t\rightarrow\infty}h^{\prime}=0$, e.g.
the solution is an eigenvector of the flux Jacobian $D\mathbf{j}$ with zero
eigenvalue. Consequently, the $(Dj)_{t\rightarrow\infty}=(D\mathbf{j}%
)(\mathbf{u}_{stat})$ is a matrix with zero eigenvalue, i.e. $\mathbf{u}%
_{stat}$ belongs to a subregion $G_{Y0X}$ with zero characteristic speed,
situated "in between" $G_{Y}$-type and $G_{X}$-type states. Such a subregion
is the boundary between $G_{Y}$-type and $G_{X}$-type domains, a hyperplane of
dimension $K-1$ characterized by $c_{p}=0$. The respective rarefaction wave is
called $p$-rarefaction wave \cite{Lax73},\cite{Lax2006}.

Arguments presented above and in
Sec.\ref{sec::Shock waves interaction as a mechanism of the phase transitions of the first order}
imply a number of consequences for the locations of continuous and
discontinuous BDPTs, discussed below.

Note that the scenario of a rarefaction wave governing long-time evolution may
take place only if initial states $G_{Y}$ on the left (and $G_{X}$ on the
right) are supported by respective boundaries, meaning that left (right)
boundary density is of $G_{Y}$ -type (of $G_{X}$ type). Such a setting appears
along a Ph2 path, see Sec.\ref{sec::A minimal path}, and never appears along a
Ph1 path. Inspecting a Ph2 path one finds that such a setting appears in the
intermediate part of the Ph2 path marked by dashed line in
Fig.\ref{Fig_MinPath}, starting as soon as the left boundary density crosses
the $c_{p}=0$ hyperplane (during step L) and finishing when the right boundary
density crosses the $c_{p}=0$ hyperplane (during step R). All along this
intermediate Ph2 segment, the rarefaction wave governs the stationary state
which stays "pinned" to the $c_{p}=0$ hyperplane. Initial and final points of
the segment are points where the pinning and depinning from the $c_{p}=0$
hyperplane take place (see also Table 2). This conclusion is fully supported
by numerical simulations. \begin{figure}[h]
\centerline{{\includegraphics[width=6.cm,height=5.5cm,clip]{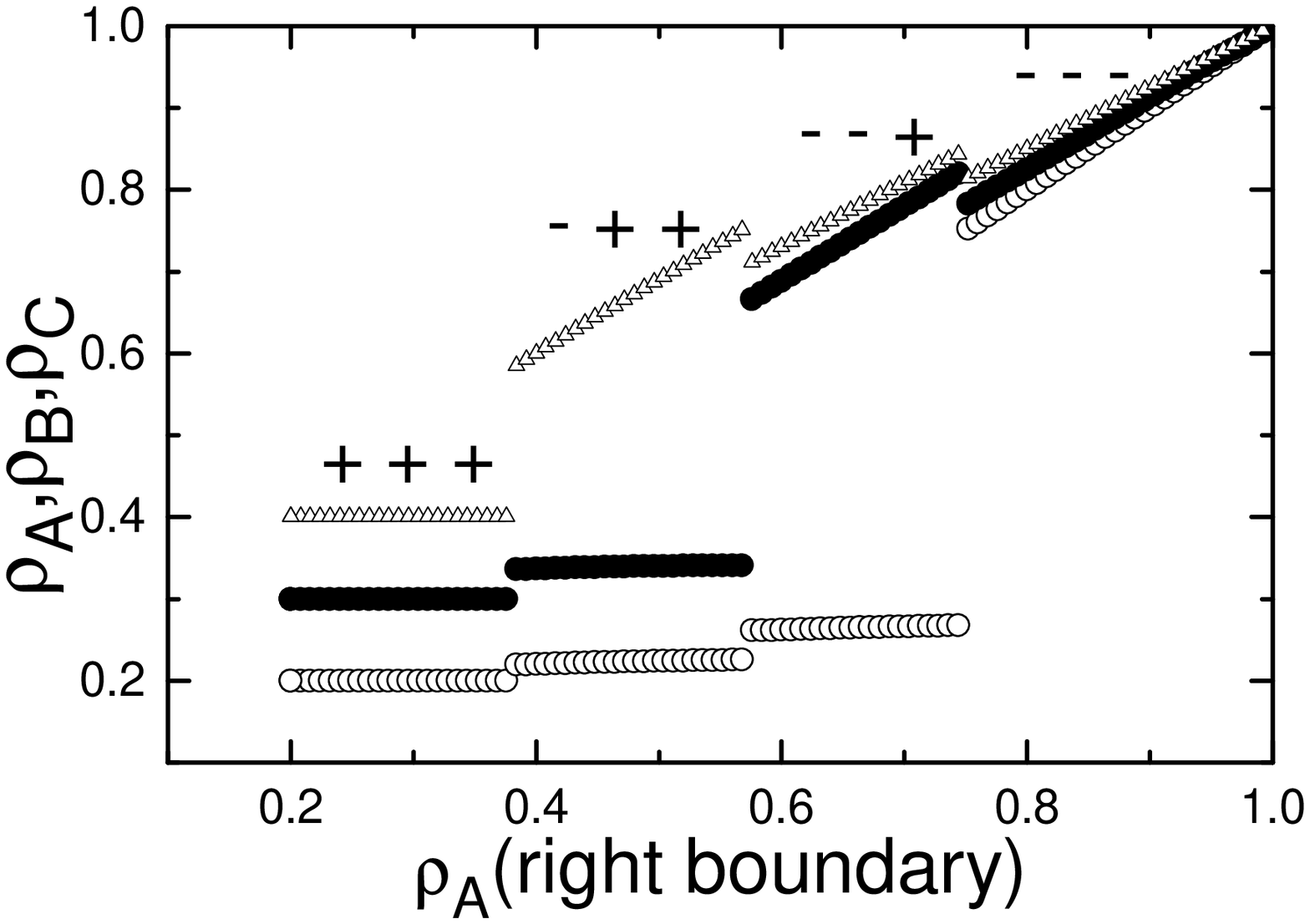}}
}\caption{Stationary densities $\rho_{A},\rho_{B},,\rho_{C}$, along a Ph1 path
across domains $G_{+++}\rightarrow G_{---}$, versus running coordinate $s$
along the path, represented by right boundary density of the first specie A
along the path) for a three-chain model $K=3$, in the torus setting where each
chain have two chains- neighbours, see \cite{Mario_multiASEP}. A particle hops
to the right neighbouring site with rate $r_{n}=1-n\gamma/4$, where $0\leq
n\leq4$ is the number of particles on the adjacent chains, neighbouring to the
departure and to the target sites. Parameters: $\gamma=0.5$. Initial and final
points of the path are \textbf{INI}$=(0.2,0.3,0.4)$ and \textbf{FIN}
$=(1,1,1)$. Three discontinuous transitions, between the states $G_{+++}%
\rightarrow G_{-++}\rightarrow G_{--+}\rightarrow G_{---}$ are clearly seen. }%
\label{Fig_K3}%
\end{figure}

Analogously, a shock wave leading to the discontinuous phase transition is
stable only if the left and the right boundary are of $G_{X}$- and of $G_{Y}%
$-type (\ref{G domains}) respectively. Such a setting always appears along a
Ph1 path (the segment marked by bold dotted line in Fig.\ref{Fig_MinPath}).
However, for an existence of a stable unbiased shock, other conditions must be
fulfilled, namely: (i) perfect balance between particle currents at both sides
of discontinuity (ii) shock densities at both sides of discontinuity must form
stable boundary layers with respective boundaries, i.e. to belong to
respective reflection maps of $u_{L}$ and $u_{R}$ \cite{reflections_JPA}.

Since the latter maps depend on the microscopic details of the dynamic, see
\cite{reflections_JPA},\cite{reflections_ABO}, we cannot locate precisely the
phase transition point, but deduce that it must be inside the unbiased
shock-wave favourable segment marked by bold dotted line in
Fig.\ref{Fig_MinPath}.

On the other hand, the unbiased shock-wave favourable setting never appears
along a Ph2 path. Therefore, discontinuous changes in stationary densities
described by our shock wave scenario, cannot happen along Ph2 path.
Consequently, \textit{ any Ph2 path }$\rho_{stat}(s)$ \textit{in physical
region is always continuous}, see Fig. \ref{Fig_MinPath} (b). Reciprocally, a
favourable setting for stable rarefaction wave formation never appears along a
Ph1 path. Therefore, a state with $c_{p}=0$, governed by a stable rarefaction
wave, cannot be observed along any Ph1 path. \begin{figure}[h]
\centerline{{\includegraphics[width=5.5cm,height=5.5cm,clip]{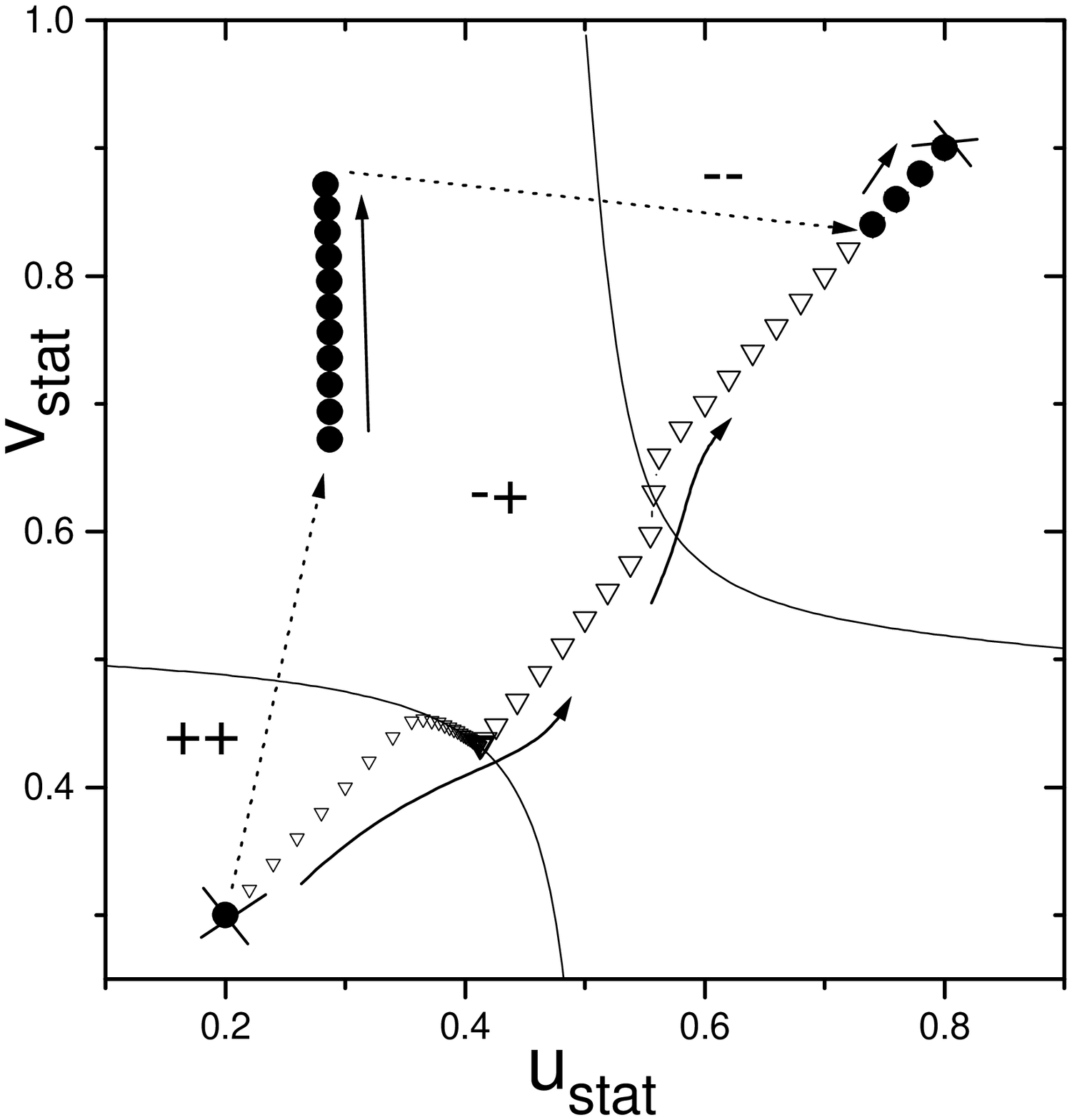}}
}\caption{Location of stationary densities along the single Ph1 path (filled
circles) and single Ph2 path (open triangles) from $G_{++}$ to $G_{--}$
domain, for two-chain model with $\gamma=0.5$. Evolution direction is marked
by arrows. Crosses show the initial and final points.}%
\label{Fig_IniFinalAll}%
\end{figure}Consequently, since initial and final stationary states
$\rho_{INI}$ and $\rho_{FIN}$ belong to different regions with $c_{p}>0$ and
$c_{p}<0$, \textit{at least one discontinuous change must happen along any Ph1
path,} see Fig. \ref{Fig_MinPath} (b).

It should be clear from our reasoning that one can construct other, more
complicated paths in parameter space, along which one can observe the same
phenomenon of discontinuous or continuous phase transitions. Any path in
parameter space connecting points $\rho_{INI}$ and $\rho_{FIN}$ in different
$G$-regions, and not containing segments favouring rarefaction waves, will
result in discontinuous phase transitions in physical region (i.e. will be
Ph1-like). Reciprocally, any path not containing segments favouring
shock-waves, (and containing therefore favorable boundary settings for stable
rarefaction waves formation, will show only continuous phase transitions (i.e.
will be Ph2 -like). Further examples are given below.

\section{Special paths for sequences of BDPTs}

\label{sec::Special paths}

Special sequences of phase transitions in system with $K$ species can be
observed along rather simple paths.

\textbf{i) Ph1 and Ph2 paths between disjoint }$G_{X}$\textbf{ and }$G_{Y}%
$\textbf{ domains.}

\begin{figure}[ptbh]
\centerline{ \subfigure[\label{Fig_MCPh1}]
{\includegraphics[width=5.5cm,height=5.5cm,clip]{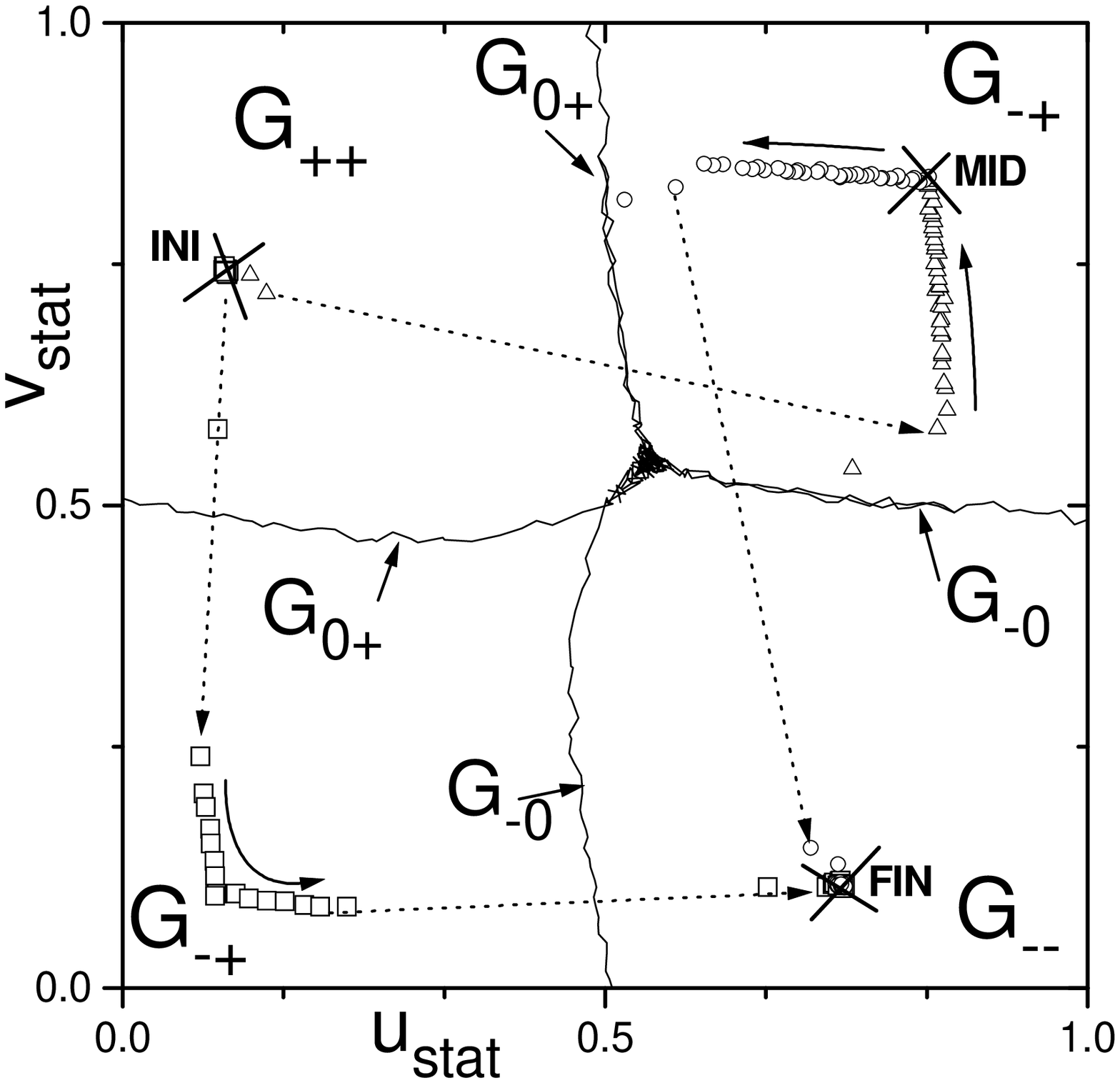}}}
\centerline{ \subfigure[\label{Fig_MCPh2}]
{\includegraphics[width=5.5cm,height=5.5cm,clip]{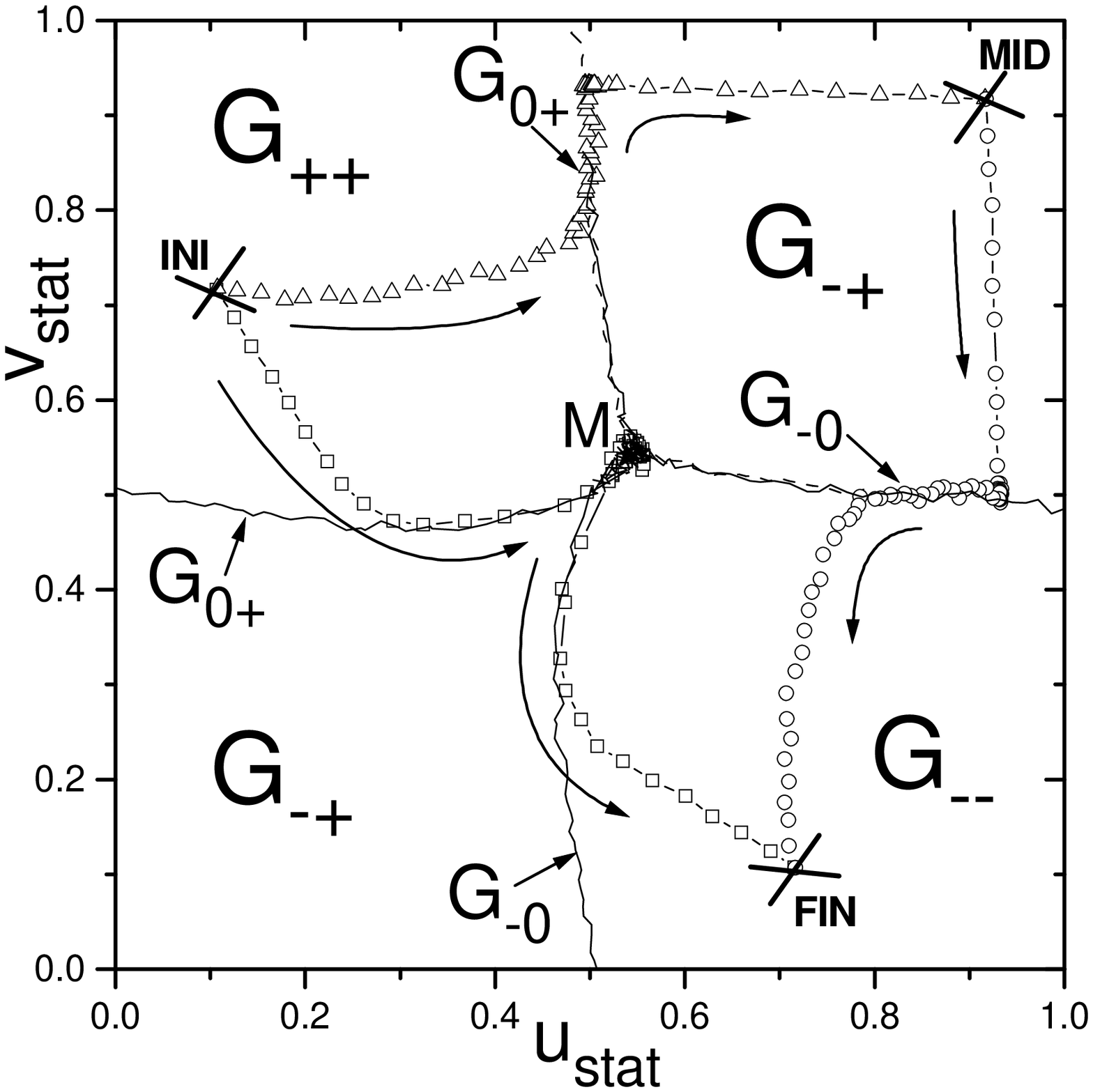}} } \caption{
Stationary densities of right and left-movers $(u_{stat},v_{stat})\equiv
\rho_{stat}$ for bidirectional traffic model from Monte Carlo simulations for
a system with $300$ sites, along various Ph1 paths (\textbf{Panel (a)}) and
Ph2 paths (\textbf{Panel (b)}). Reference Initial, Middle and Final points are
marked by crosses \textbf{INI, MID } and \textbf{FIN}). Lines where one
characteristic velocity is zero ($G_{0+},G_{-0}$) are obtained numerically.
Evolution of the $\rho_{stat}(s)$ along a path is marked by arrows, dotted
arrows mark discontinuous transitions. Few data points outside the
arrow-marked paths result from finite size effects. The symmetry of the Figure
with respect to the line $y=x$ is due to the left-right symmetry of the model
and the points \textbf{INI, FIN }. Parameters: $h=0.5$. \textbf{Panel (a)}:
Squares mark $\rho_{stat}$ along a Ph1 path which goes directly from the
initial to the final point \textbf{INI}$\rightarrow$ \textbf{FIN}. Triangles
and circles mark $\rho_{stat}$ along two consecutive Ph1 paths \textbf{INI}%
$\rightarrow$ \textbf{MID}, \textbf{MID}$\rightarrow$ \textbf{FIN}. Initial,
Middle and Final boundary rates, corresponding to points \textbf{INI,MID,FIN}
are $(\alpha=1-\beta=0.1,A=1-B=0.78)$; $(\alpha=1-\beta=A=1-B=0.9)$ and
$(A=1-B=0.1,\alpha=1-\beta=0.78)$. Along all paths, the boundary rates
$\alpha,\beta,A,B$ are changed by linear interpolation law. \textbf{Panel
(b)}: The same as Panel (a), for respective Ph2 paths. Note that the densities
along the direct Ph2 path (squares) go through the weak hyperbolic point
$c_{1}=c_{2}=0$, marked by $M$, see also discussion at the end of
Sec.\ref{sec::Special paths}. Initial,Middle and Final boundary rates:
($\alpha=1-\beta=0.1,A=1-B=0.75$), $(\alpha=1-\beta=A=1-B=0.95)$, and
$(A=1-B=0.1,\alpha=1-\beta=0.75)$ respectively.}%
\label{Fig_MC}%
\end{figure}

Along any \emph{single} (not composite) Ph1-like path (\ref{Ph1}) connecting
arbitrary \textit{disjoint} $G$ regions, a sequence of first order phase
transitions will be observed, provided that the initial state has more
positive characteristic speeds than the final state. The latter condition
makes existence of stable shocks (governing first order transitions) possible
and leads to observation of as many discontinuous transitions, as the number
of hyperplanes $c_{p}=0$ separating the initial and the final state. The
existence of such a path was pointed out in \cite{PopkovCambridge}. For
example, to observe all qualitatively different first order transitions, we
can take a single Ph1 path (\ref{Ph1}) from initial point with all positive
characteristics $\rho_{ini}\in G_{+..+}$ to a final point with all negative
characteristics $\rho_{final}\in G_{-...-}$. Along such a path, $K$ first
order transitions will be observed, see Figs.\ref{Fig_IniFinalAll}%
,\ref{Fig_K3}, and the path marked by squares in
Fig.\ref{Fig_MC}(a). In particular, Fig.\ref{Fig_K3} corresponds
to multi-chain model with $\ K=3$ and shows respectively three
discontinuous transitions in stationary density along the Ph1
path. The model has product stationary states which allows to
compute the particle fluxes and consequently characteristic
velocities analytically as functions of particle densities.

By computing the characteristic velocities along the path in
physical region $\rho_{stat}(s)$ for $K=3$ we find that across each
discontinuous transition just one characteristic velocity changes sign, this
confirming the first-order phase transition scenario described in Sec.
\ref{sec::Shock waves interaction as a mechanism of the phase transitions of the first order}%
.

Similarly, a \textit{single} Ph2 path from initial to final state which belong
to disjoint $G$ regions allows to observe the sequence of all continuous
transitions between these states. As examples, see Figs. \ref{Fig_K3},
\ref{Fig_IniFinalAll} (see also the path marked by squares in Fig.\ref{Fig_MC}%
(b) of the next section for a more complex model). Note that it is important
that that the initial state has more positive characteristic speeds than the
final state.

\noindent\textbf{ii) Fully matching path.}

Another special path is a path where left and right boundary densities are
equal all along from the initial till the end path point. It is clear that in
this case we will not observe any phase transitions because there will be
always a perfect match of the bulk density with the boundaries. Consequently,
this path in parameter space must contain all triple points where the
hyperplanes of second order and first order transitions merge together. For
$K=1$ such a triple point is a point ($\rho^{\ast},\rho^{\ast}$) where the
characteristic speed vanishes $j^{\prime}(\rho^{\ast})=0$. The nature and
topology of the parameter space in vicinity of these triple points (lines,
hypersurfaces) will be discussed elsewhere.

\section{Model for a bidirectional traffic on a narrow road}

\label{sec::Two way traffic model} In the previous sections we concentrated
our attention on solvable models with analytic flux functions and strict
hyperbolicity. For generic (not integrable) models, however, the analytic flux
function is typically unknown, as well as\ \ exact relation between boundary
rates and effective reservoir boundary densities. In the following we show how
even in this case it is still possible to construct Ph1-like or Ph2-like paths
along which a given BDPT type (discontinuous or continuous) can be observed.
As an example, we consider the case of a two-way traffic model on a narrow
road (see bottom panel of Fig. 3).

Models of bidirectional traffic have been widely studied in the literature and
appear in several contexts see e.g. \cite{Juhasz-Bidirectional2010}. Our
system consists of two chains, containing particles hopping in opposite
direction: a particle hops in preferred direction with constant rate $1$ and
hard core exclusion like in TASEP, but is slowing down when it meets an
upcoming particle (an obstacle) in front on the adjacent lane: in this case
the rate of hopping is $\exp(-h)$, where a positive constant $h$ measures the
interlane interaction, see Fig.\ref{Fig_TwoWayProcesses}. A similar model, but
with periodic boundary conditions, was considered in
\cite{Jiang-Bidirectional09}. We choose the boundary rates as follows: if the
target site is vacant, a right-moving particle can enter with rate $\alpha$
($\alpha e^{-h}$) if the adjacent to the target site is empty (is occupied by
an upcoming particle). At the other end, a particle can leave with rate
$\beta$. For the left moving particles, the entrance and exit rates are
respectively $A$($Ae^{-h}$) and $B$. Note, that the model has the left-right
symmetry. Since the model is not solvable, the analytical expression for the
flux $j(u,v,h)$ is not known for any nonzero $h$. Neither we know the exact
relation between the boundary rates and the effective boundary densities.

Ph1- and Ph2-like paths, however, can be constructed straightforwardly. From
the physical meaning of the characteristic velocities (e.g. velocities with
which small perturbations of the homogeneous state propagate)
\cite{GunterSlava_StatPhys} we conclude that a stationary state with small
density of right moving particles or right moving holes realized e.g. for
$\alpha=1-\beta\ll1$ and $B=1-A\ll1$, has all positive characteristic
velocities and therefore it must be in the $G_{++}$ region. By left-right
symmetry, a stationary state with small density of left moving particles or
holes will belong to the $G_{--}$ region ( the respective boundary rates are
attainable from $G_{++}$ rates by exchanging $\alpha\Longleftrightarrow
A,\beta\Longleftrightarrow B$). Finally, a state with small density of right
movers on one lane and small density of left movers on another lane, realized
by $\alpha,A,1-\beta,1-B\ll1$ or $1-\alpha,1-A,\beta,B\ll1$, belongs the
$G_{-+}$ region. Proceeding along Ph1 (Ph2) paths in parameter space between
regions $G_{++}\rightarrow$ $G_{-+}$, and $G_{-+}\rightarrow G_{--}$, one
expects to see the occurrence of first (second) order phase transitions as
described above. This is precisely what we obtain from Monte Carlo simulations
of the two-way model, see Fig.\ref{Fig_MC}.

We can also build direct Ph1 and Ph2 paths between $G_{++}\rightarrow$
$G_{--}$ as described in Sec.\ref{sec::Special paths} i, see square data
points in Fig.\ref{Fig_MC}. Note that along a direct Ph2 path (see
Fig.\ref{Fig_MC}(b)), the pinning/depinning of stationary densities to the
line with $c_{p}=0$ occurs only once, due to a presence of a special point (or
region) $M$ in the middle where the lines $c_{1}=0$ and $c_{2}=0$ intersect.
Such a point where two characteristic velocities coincide (the so- called
weakly hyperbolic point), makes possible a continuous passage from $G_{++}$ to
$G_{--}$ domain. It is worth to note that according to the numerical study of
Jiang et al. \cite{Jiang-Bidirectional09} restricted to the case of periodic
boundary conditions and equal particle densities, the steady state current
along the symmetric line $u_{stat}=v_{stat}$ develops a plateau, leading in
periodic system to phase separation. Such a non-analiticity in the stationary
current suggests that the region $M$ in the middle of Fig.\ref{Fig_MC}(b) is a
segment rather that a single point.

It is quite remarkable that even in this, rather special situation with
non-analytic current-density dependence, our predictions about
discontinuity/continuity of phase transitions along Ph1/Ph2 paths remain
robust. We also remark that the presence of the region M can be neglected as
long as our paths are situated far enough from it. The systematic study of an
influence of a weakly hyperbolic point on BDPTs will be done elsewhere.

\section{Conclusions}

\label{sec::Conclusions}

In this paper we have classified the basic phase transitions which can be
observed in multi-species driven systems with open boundaries. We have shown
that the splitting of the physical region into domains with different signs of
characteristic speeds, and hyper-surfaces separating these regions where one
of characteristic speeds vanishes, plays a fundamental role in this
classification. Adiabatic paths in the parameter space, defined by the
particle densities of each specie at the left and right boundary reservoirs,
along which we surely observe discontinuous or continuous transition of a
desired type, or a desired sequence of BDPTs, have been explicitly
constructed. The details of the microscopic dynamics and the geometry of the
models are not important for our qualitative BDPTs scenarios to occur, as far
as several conditions listed at the beginning of
Sec.\ref{sec::Multi-species particle models out of equilibrium} are fulfilled.
We expect therefore our results to be valid for a broad class of particle
models with several interacting particle species. In particular, our examples
were systems of particles obeying hard-core exclusion rule, but this is not
required as far as some interaction making the flux function nonlinear will be
present.

Mathematically, our study has been focused mainly to models with analytic flux
function and strict hyperbolicity i.e. models with Jacobian matrices which
have distinct eigenvalues in all the physical region. An example of a weakly
hyperbolic model with non-analytic flux and phase separation, however, was
considered in Sec.\ref{sec::Two way traffic model}. It is remarkable that even
for this model the general validity of our approach has been confirmed. An
interesting problem for the future would be to test the predictions of our
analysis on more complicated models, like those showing symmetry breaking,
hysteresis and ergodicity breaking phenomena.

\section{Acknowledments}

It is a pleasure to thank G. Schütz for discussion and valuable comments. VP
wish to thank the Department of Physics and the University of Salerno, for
hospitality and for providing a research grant (Assegno di Ricerca no. 1508,
2007-2010) during which this work was done. This work has been partially
supported by the DFG grant 436 RUS 113/909/0-1(R) and by the Italian Ministry
for Education, University and Research (MIUR) through an inter-University
PRIN-2008 initiative.

%\bibliography{C:/Programmi/swp50/TCITex/BibTeX/bib/ABO}

\begin{thebibliography}{999999999999999999999999999999999999999999999999999999999999999999}                               %


\expandafter\ifx\csname natexlab\endcsname\relax
\fi
\expandafter\ifx\csname bibnamefont\endcsname\relax


\fi
\expandafter\ifx\csname bibfnamefont\endcsname\relax


\fi
\expandafter\ifx\csname citenamefont\endcsname\relax


\fi
\expandafter\ifx\csname url\endcsname\relax


\fi
\expandafter\ifx\csname urlprefix\endcsname\relax


\fi
\providecommand{\bibinfo}[2]{#2} \providecommand{\eprint}[2][]{\url{#2}}

%\cite{Gunter03TwoSpecies_review},\cite{KoloFischer07_review}%
%,\cite{Frey09}


\bibitem[Gunter(03)]{Gunter03TwoSpecies_review}%
\bibinfo{author}{\bibfnamefont{G.}~\bibnamefont{Sch\"utz}},
\bibinfo{journal}{J. Phys. A} \textbf{\bibinfo{volume}{36}},
\bibinfo{pages}{R339} (\bibinfo{year}{2003}).

\bibitem[KoloFischer(2007)]{KoloFischer07_review}%
\bibinfo{author}{\bibfnamefont{A.B.~Kolomeisky} \bibfnamefont{and} \bibnamefont{M.E.~Fisher}},
\bibinfo{journal}{Annu. Rev. Phys. Chem.} \textbf{\bibinfo{volume}{58}},
\bibinfo{pages}{675} (\bibinfo{year}{2007}).

\bibitem[3]{Frey09}%
\bibinfo{author}{\bibfnamefont{A.~Basu} \bibfnamefont{and} \bibnamefont{E.~Frey}},
\bibinfo{journal}{J. Stat. Mech.}
%\textbf{\bibinfo{volume}{36}}
, \bibinfo{pages}{p. P09013} (\bibinfo{year}{2009}).

\bibitem[Krug(1991)]{Krug91}%
\bibinfo{author}{\bibfnamefont{J.}~\bibnamefont{Krug}},
\bibinfo{journal}{Phys.
Rev. Lett.} \textbf{\bibinfo{volume}{67}}, \bibinfo{pages}{1882} (\bibinfo{year}{1991}).

\bibitem[Kolomeisky et~al.(1998)Kolomeisky, Schütz, Kolomeisky, and
Straley]{Kolo98}%
\bibinfo{author}{\bibfnamefont{A.~B.} \bibnamefont{Kolomeisky}},
\bibinfo{author}{\bibfnamefont{G.~M.} \bibnamefont{Schütz}},
\bibinfo{author}{\bibfnamefont{E.~B.} \bibnamefont{Kolomeisky}}, and
\bibinfo{author}{\bibfnamefont{J.~P.}
\bibnamefont{Straley}}, \bibinfo{journal}{J. Phys. A}
\textbf{\bibinfo{volume}{31}}, \bibinfo{pages}{6911} (\bibinfo{year}{1998}).

\bibitem[Popkov(2007)]{PopkovCambridge}%
\bibinfo{author}{\bibfnamefont{V.}~\bibnamefont{Popkov}},
\bibinfo{journal}{Journal of Stat. Mechanics: Theory and Experiment} p.
\bibinfo{pages}{P07003} (\bibinfo{year}{2007}).

\bibitem[Lax(1973)]{Lax73}%
\bibinfo{author}{\bibfnamefont{P.~D.} \bibnamefont{Lax}},
\emph{\bibinfo{title}{Hyperbolic Systems of Conservation Laws and the
Mathematical Theory of Shock Waves}} (\bibinfo{publisher}{SIAM series,
Philadelphia, vol. 11}, \bibinfo{year}{1973}).

\bibitem[Lax(2006)]{Lax2006}%
\bibinfo{author}{\bibfnamefont{P.~D.} \bibnamefont{Lax}},
\emph{\bibinfo{title}{Hyperbolic Partial Differential Equations}}
(\bibinfo{publisher}{Courant Lecture Notes in Mathematics, vol. 14, New
York}, \bibinfo{year}{2006}).

\bibitem[Bressan(2000)]{Bressan}%
\bibinfo{author}{\bibfnamefont{A.}~\bibnamefont{Bressan}},
\emph{\bibinfo{title}{Hyperbolic Systems of Conservation Laws}}
(\bibinfo{publisher}{Oxford University Press, New York}, \bibinfo{year}{2000}).

\bibitem[Popkov and Schütz(1999)]{Gunter_Slava_Europhys}%
\bibinfo{author}{\bibfnamefont{V.}~\bibnamefont{Popkov}} and
\bibinfo{author}{\bibfnamefont{G.~M.} \bibnamefont{Schütz}},
\bibinfo{journal}{Europhys. Lett} \textbf{\bibinfo{volume}{48}},
\bibinfo{pages}{257} (\bibinfo{year}{1999}).

\bibitem[Popkov and Sch{\"u}tz(2004)]{reflections_ABO}%
\bibinfo{author}{\bibfnamefont{V.}~\bibnamefont{Popkov}} and
\bibinfo{author}{\bibfnamefont{G.~M.} \bibnamefont{Sch{\"u}tz}},
\bibinfo{journal}{J.Stat. Mech.:Theory and Experiment} p.
\bibinfo{pages}{P12004} (\bibinfo{year}{2004}).

\bibitem[Wea()]{Weak_Hyperbolicity}%
\bibinfo{note}{For some models, the characteristic velocities may coincide at
certain parameter values. This leads in the hydrodynamic limit to
weakly hyperbolic equations, and extra phenomena, see e.g.
\cite{Peschel}. Here we focus on models with strict hyperbolicity,
where the characteristic velocities are always distinct.}

\bibitem[Popkov and Schütz(2003)]{GunterSlava_StatPhys}%
\bibinfo{author}{\bibfnamefont{V.}~\bibnamefont{Popkov}} and
\bibinfo{author}{\bibfnamefont{G.~M.} \bibnamefont{Schütz}},
\bibinfo{journal}{J Stat. Phys.} \textbf{\bibinfo{volume}{112}},
\bibinfo{pages}{523} (\bibinfo{year}{2003}).

\bibitem[Schütz(2000)]{Schu00}%
\bibinfo{author}{\bibfnamefont{G.~M.} \bibnamefont{Schütz}},
\emph{\bibinfo{title}{Exactly solvable models for many-body systems far from
equilibrium}} (\bibinfo{publisher}{Academic Press, London},
\bibinfo{year}{2000}), \bibinfo{note}{in: Phase Transitions and Critical
Phenomena, ed. C.Domb and J.L. Lebowitz, Vol. 19}.

\bibitem[Derrida(1998)]{TASEP_reviewDerrida}%
\bibinfo{author}{\bibfnamefont{B.}~\bibnamefont{Derrida}},
\bibinfo{journal}{Physics Reports} \textbf{\bibinfo{volume}{301}},
\bibinfo{pages}{65} (\bibinfo{year}{1998}).

\bibitem[Popkov and Salerno(2004)]{Mario_multiASEP}%
\bibinfo{author}{\bibfnamefont{V.}~\bibnamefont{Popkov}} and
\bibinfo{author}{\bibfnamefont{M.}~\bibnamefont{Salerno}},
\bibinfo{journal}{Phys. Rev. E} \textbf{\bibinfo{volume}{69}},
\bibinfo{pages}{046103} (\bibinfo{year}{2004}).

\bibitem[Popkov(2004)]{reflections_JPA}%
\bibinfo{author}{\bibfnamefont{V.}~\bibnamefont{Popkov}},
\bibinfo{journal}{J.
Phys. A} \textbf{\bibinfo{volume}{37}}, \bibinfo{pages}{1545} (\bibinfo{year}{2004}).

\bibitem[Juhasz(2010)]{Juhasz-Bidirectional2010}%
\bibinfo{author}{\bibfnamefont{R.}~\bibnamefont{Juhasz}},
\bibinfo{journal}{J
Stat Mech} p. \bibinfo{pages}{P03010} (\bibinfo{year}{2010}).

\bibitem[Jiang et~al.(2009)Jiang, Nishinari, Hu, Wu, and Wu]%
{Jiang-Bidirectional09}%
\bibinfo{author}{\bibfnamefont{R.}~\bibnamefont{Jiang}},
\bibinfo{author}{\bibfnamefont{K.}~\bibnamefont{Nishinari}},
\bibinfo{author}{\bibfnamefont{M.~B.} \bibnamefont{Hu}},
\bibinfo{author}{\bibfnamefont{Y.~H.} \bibnamefont{Wu}}, and
\bibinfo{author}{\bibfnamefont{Q.~S.} \bibnamefont{Wu}}, \bibinfo{journal}{J
Stat Phys} \textbf{\bibinfo{volume}{136}}, \bibinfo{pages}{73} (\bibinfo{year}{2009}).

\bibitem[Popkov and Peschel(2001)]{Peschel}%
\bibinfo{author}{\bibfnamefont{V.}~\bibnamefont{Popkov}} and
\bibinfo{author}{\bibfnamefont{I.}~\bibnamefont{Peschel}},
\bibinfo{journal}{Phys. Rev. E} \textbf{\bibinfo{volume}{64}},
\bibinfo{pages}{026126} (\bibinfo{year}{2001}).
\end{thebibliography}

\end{document}